\documentclass[12pt]{iopart}
\usepackage{graphicx}
\usepackage{color}
\usepackage{setstack}
\usepackage{amstext}
\usepackage{amssymb}
\usepackage{fixltx2e}

\definecolor{orange}{rgb}{1,0.5,0}

\begin{document}
\title[2D capillary wetting]{Wetting in a two-dimensional capped capillary. Part II: Three-phase coexistence.}
\author{P Yatsyshin$^1$ and N Savva$^{2}$ and S Kalliadasis$^1$}
\address{$^1$ Department of Chemical Engineering, Imperial College London,
London SW7 2AZ, United Kingdom}
\address{$^2$ School of Mathematics, Cardiff University, Cardiff, CF24 4AG, United Kingdom}
\ead{\mailto{p.yatsyshin@imperial.ac.uk},\mailto{SavvaN@cardiff.ac.uk},
\mailto{s.kalliadasis@imperial.ac.uk}}
\begin{abstract}
In Part II of this study we consider two cases of three-phase coexistence.
First, the capped capillary may allow for vapour, drop-like, and slab-like
phases to coexist at the same values of temperature and chemical potential.
Second, the slit pore forming the bulk of the capped capillary may allow for
the coexistence between vapour, planar prewetting film and capillary-liquid.
While the consideration of the former case allows us to summarise the
phenomenology presented in Part I and to show that the transition line of
wedge prewetting is shifted in capillary-like geometries by a constant value,
depending on the capillary width, the careful examination of the latter case
allows us to uncover a new phase transition in confined fluids, a continuous
planar prewetting transition. A planar prewetting transition is known to be a
distinctly first-order phenomenon, and typically taking place on the scale of
several atomic diameters. A continuous prewetting transition, on the other
hand, is scale invariant. Thus, apart from being of fundamental significance,
this finding has potential for facilitating experimental detection as well as
measurements of planar prewetting. Further, we provide proof for the
existence of a tri-critical point of the three-phase coexistence line of the
capped capillary while by considering a dynamic model of wetting we show how
the relaxation of the system can be pinned by a metastable state. We present
a full parametric study of our model system and support our findings with
exhaustive examples of density profiles, adsorption and free energy
isotherms, and full phase diagrams.

\end{abstract}
\submitto{\JPCM}
\maketitle
\section{Introduction}

In Part I of this study we have shown how capping a one-dimensional (1D) slit
pore, which results in a generically two-dimensional (2D) system,
dramatically changes the phenomenology of wetting. The 2D system possesses
the capillary wetting temperature, $T_{\text{cw}}$, which is the property of
the pore and controls the order of capillary condensation (CC) at any given
value of temperature. We have also shown the existence of capillary
prewetting transition, where configurations of the pore filled with vapour
coexist with those possessing a finite capillary-liquid slab. Finally, we
have shown that wedge wetting can manifest itself in the capped capillary by
the existence of a (metastable) fluid phase with drops in the capillary
corners. The details of the density functional (DF) model, working equations
and discussion of the numerical methodology can be found in Part I and we
will be referring to them here when necessary. As the wetting behaviour is
qualitatively the same, whether repulsions are treated with a non-local
(weighted density approximation, WDA) or a local (local density
approximation, LDA) functional, for computational convenience we adopt the
latter approximation here.

To complete the description of phase behaviour of fluids confined in capped
capillaries we focus on the (most general) case which exhibits two types of
three-phase coexistence. First, the 1D capillary bulk, given by the
associated slit pore, has three coexisting phases: vapour, capillary-liquid
and planar thin film adsorbed on each wall. Second, the 2D capped capillary
exhibits the coexistence of surface phases of vapour, capillary-liquid slab
and capillary-liquid drops in the corners. The interplay of prewetting with
continuous CC leads to the existence of a new type of phase transition --
continuous prewetting. We proceed to addressing all details of three-phase
coexistence and we shall also areas, where further research, preferably going
beyond mean-field and including fluctuation effects, will be necessary.
\begin{figure}
\centering
    \includegraphics{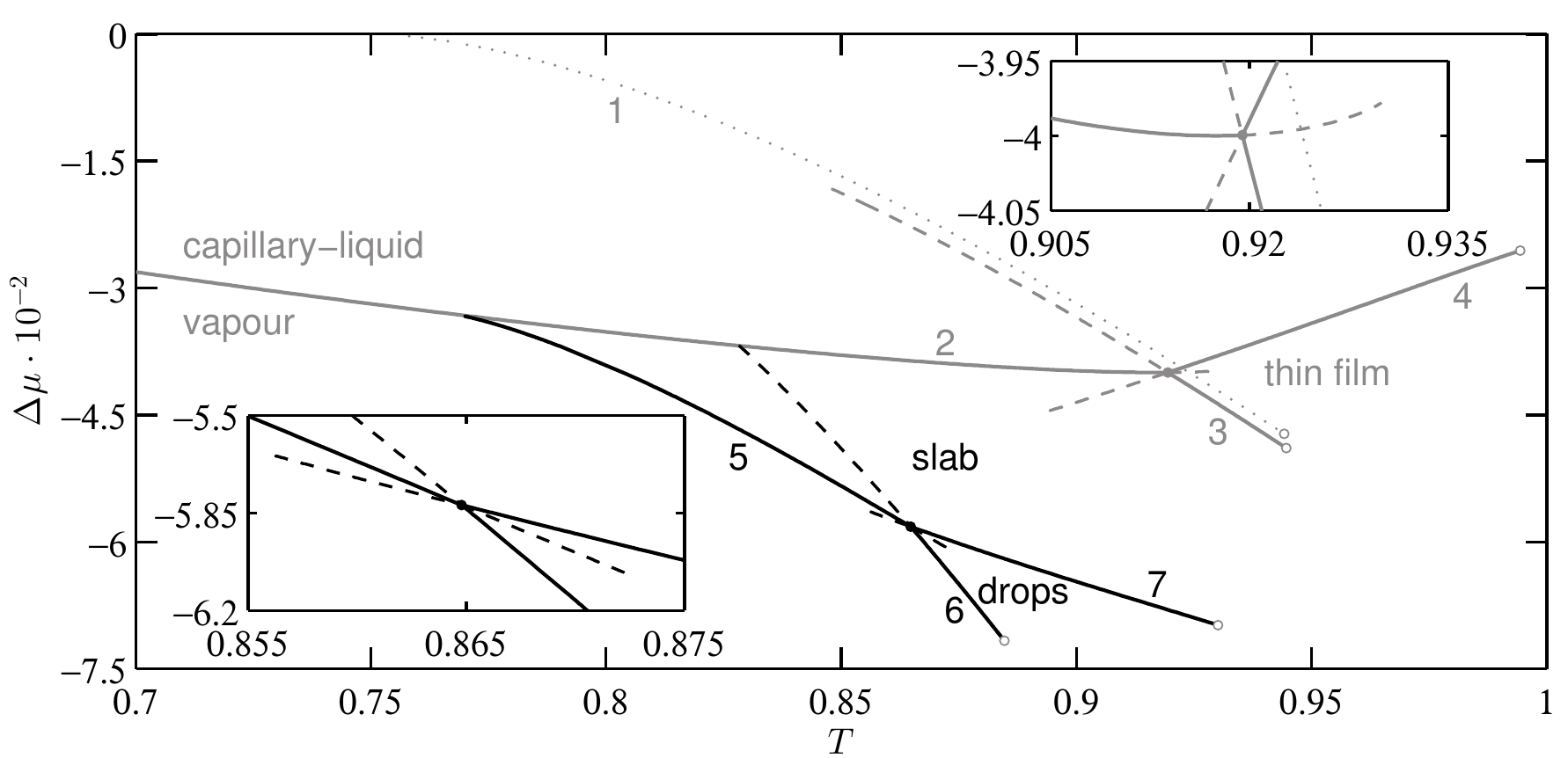}%
    \caption{Phase diagram of the capped capillary with $\varepsilon_{\text{w}}=0.7$,
    $\sigma_{\text{w}}=2$, $H_0 = 5$, $H=40$; fluid treated in LDA, planar $T_{\text{w}} = 0.755$.
    The system exhibits all effects discussed in Part I. Grey color pertains to the capillary bulk
    (slit pore associated with the capped capillary), black color to the capped capillary.
    Regions where a single phase is thermodynamically stable are labelled by a text note.
    Each transition line is numbered and consists of two parts (except the planar prewetting line (1, grey, dotted)):
    solid part denotes a transition between thermodynamically stable phases, dashed part between metastable phases.
    Filled circles denote triple points. Open circles denote critical points. Insets zoom on the regions containing the
    triple points of the slit pore, $T_3^{\text{slt}}=0.920$ and of the capped capillary, $T_{3}^{\text{cpd}}=0.865$.
    Note that the CC transition line consists of stable parts of lines 2 and 4, and capillary prewetting line of stable parts of lines 5 and 7.

    List of transition lines by their labels, with a brief description of coexisting phases.

    1: planar prewetting ($\Delta\mu_{\text{pw}}\left(T\right)$, vapour transforms to thin film), begins at $\left(T_{\text{w}},0\right)=\left(0.755,0\right)$, ends at $\left(T_{\text{pw}}^{\text{cr}},\Delta\mu_{\text{pw}}^{\text{cr}}\right)=\left(0.944,-4.72\cdot10^{-2}\right)$,

    2 and 4: CC ($\Delta\mu_{\text{cc}}\left(T\right)$, transition to capillary-liquid from vapour (line 2) and from thin film (line 4)), line 2 ends at $\left(0.929,-3.98\cdot10^{-2}\right)$, line 4 begins at $\left(0.89,-4.45\cdot10^{-2}\right)$, ends at $\left(T_{\text{cc}}^{\text{cr}},\Delta\mu_{\text{cc}}^{\text{cr}}\right)=\left(0.994,-2.56\cdot10^{-2}\right)$,

    3: shifted planar prewetting ($\Delta\tilde{\mu}_{\text{pw}}\left(T\right)$, vapour to thin film in the associated slit pore), begins at $\left(0.848,-1.83\cdot10^{-2}\right)$, ends at $\left(\tilde{T}_{\text{pw}}^{\text{cr}},\Delta\tilde{\mu}_{\text{pw}}^{\text{cr}}\right)=\left(0.945,-4.89\cdot10^{-2}\right)$,

    5 and 7: capillary prewetting ($\Delta\mu_{\text{cpw}}\left(T\right)$, transition to capillary-liquid slab from vapour (line 5) and from drop phase (line 7)), line 5 begins at $\left(T_{\text{cw}},\Delta\mu_{\text{cw}}\right)=\left(0.77,-3.33\cdot10^{-2}\right)$ ($\Delta_{\mu,{\text{cc}}}=0.0086\cdot10^{-2}$), ends at $\left(0.873,-6.1\cdot10^{-2}\right)$, line 7 begins at $\left(0.856,-5.65\cdot10^{-2}\right)$, ends at $\left(\tilde{T}_{\text{cpw}}^{\text{cr}},\Delta\tilde{\mu}_{\text{cpw}}^{\text{cr}}\right)=\left(0.93,-6.98\right)$,

    6: shifted wedge prewetting ($\Delta\tilde{\mu}_{\text{wpw}}\left(T\right)$, vapour to drops in corners), begins at $\left(\tilde{T}_0,\Delta\tilde{\mu}_0\right)=\left(0.828,-3.68\cdot10^{-2}\right)$ (point is on the CC line (line 2)), ends at $\left(\tilde{T}_{\text{wpw}}^{\text{cr}},\Delta\tilde{\mu}_{\text{wpw}}^{\text{cr}}\right)=\left(0.885,-7.17\cdot10^{-2}\right)$.
\label{Fig3}}
\end{figure}

Central to our discussion is figure \ref{Fig3}, which shows the full phase
diagram of the fluid in the capped capillary with the substrate parameters
$\varepsilon_{\text{w}}=0.7$, $\sigma_{\text{w}}=2$, $H_0 = 5$ and width
$H=40$. The fluid is treated in LDA and the planar wetting temperature is
$T_{\text{w}}=0.755$. Note that the capillary in figure \ref{Fig3} differs
from the one in figure 10 of Part I (with $H=30$) only in the value of wall
separation, $H$. The grey curves pertain to the phases inside the slit pore
(capillary bulk, a 1D system), while the black curves correspond to the
phases formed near the capping wall (capped capillary, a 2D system). Insets
zoom on the triple points. All the phase transition lines are indexed and
described in the caption of the figure. A solid line denotes transitions
between stable phases, a dashed line denotes transitions between metastable
phases. The regions of thermodynamic stability (but not metastability,
spinodals are not shown) of a single phase are clearly labelled by text
notes. Triple points are marked by filled circles. Critical points are marked
by open circles. The dotted grey line 1 is the prewetting line of a single
planar wall: the locus of phase transitions between vapour and a thin planar
film. Regarding capillary bulk (slit pore), line 2 is the locus of phase
transitions between vapour and capillary-liquid, line 3 -- between vapour and
thin film formed on each side wall, line 4 -- between thin film and
capillary-liquid. Regarding the capped capillary, line 5 is the locus of
transitions between vapour and configurations with capillary-liquid slab
(capillary prewetting), line 6 -- the locus of transitions between vapour and
configurations with drops in corners (shifted wedge prewetting), line 7 --
the locus of phase transitions between configurations with corner drops and a
capillary-liquid slab.

To obtain the full phase diagram, we had to compute two triple points for
three-phase coexistence inside the slit pore and the capped capillary. The
condition for the coexistence of three fluid phases with density
distributions $\rho_1$, $\rho_2$, $\rho_3$ can be expressed, similarly to the
condition for the two-phase coexistence, given in equation (24) of Part I, by
the requirement that each of the density profiles minimises the total free
energy (equation (11) of Part I), and that the excess free energies of all
three fluid configurations to be equal:
\begin{eqnarray}
\label{3point}
\frac{\delta\Omega}{\delta\rho}\Big |_{\rho_1}=\frac{\delta\Omega}{\delta\rho}\Big |_{\rho_2}=\frac{\delta\Omega}{\delta\rho}\Big |_{\rho_3}=0\nonumber\\
\Omega^{\text{ex}}\left[\rho_1\right]=\Omega^{\text{ex}}\left[\rho_2\right]=\Omega^{\text{ex}}\left[\rho_3\right],\end{eqnarray}
where, for a slit pore (1D system)
$\Omega^{\text{ex}}\left[\rho^{\text{slt}}\left(y\right)\right]\equiv\Omega\left[\rho^{\text{slt}}\left(y\right)\right]$,
see equation (2) of Part I, and in the case of the capped capillary (2D
system), $\Omega^{\text{ex}}\left[\rho^{\text{cpd}}\left(x,y\right)\right]$
is given in equation (21) of Part I.

To locate the triple point, one has to solve the above system for the density
profiles $\rho_1$, $\rho_2$, $\rho_3$ of the coexisting fluid configurations,
as well as for the chemical potential, $\mu_3$, and the temperature $T_3$. An
initial guess for $\mu_3$ and $T_3$ ($\mu_{3,0}$ and $T_{3,0}$) can be
obtained by interpolating the three intersecting two-phase coexistence lines
(which should intersect at a single point but due to computational error they
will intersect at three points) and then use their mean as $\mu_{3,0}$ and
$T_{3,0}$. As an initial guess for the densities one can then take the
density profiles inside each of the three phases, which correspond to the
data points on two-phase coexistence lines, closest to $\mu_{3,0}$ and
$T_{3,0}$. So the strategy is similar to finding the point of two-phase
coexistence, see equation (24) of Part I. Using arc-length continuation one
can then trace the entire triple-line in the space of, e.g., parameters of
the potential, $\varepsilon_{\text{w}}$ and $\sigma_{\text{w}}$, or wall
separation, $H$, and search numerically for a tri-critical point. We leave
the detailed numerical investigation of the triple, such as continuation with
respect to the capillary width or substrate potential parameters, for future
work.

The triple point splits each transition line passing through it into two
branches: where the transition takes place between thermodynamically stable
(solid line) and metastable (dashed line) fluid states. The stable branch of
a transition line ends at a critical point (marked by open circles in figure
\ref{Fig3}). On the other hand, the end points of metastable branches of
transition lines are not critical, they are associated with a metastable
branch of the excess free energy $\Omega^{\text{ex}}\left(\Delta\mu\right)$
losing an intersection point with a stable branch. We investigate the
mean-field signature of criticality in detail in section \ref{ThreePhaseCap}.

Considering the $\Delta\tilde{\mu}_{\text{wpw}}\left(T\right)$-lines in
figure 10 of Part I and figure \ref{Fig3}, we can, without the computation of
the full wedge prewetting line, $\Delta\mu_{\text{wpw}}\left(T\right)$,
prove, that confinement, imposed by a second side wall of the capped
capillary on the otherwise a wedge-shaped substrate, leads to the shift of
the wedge prewetting transition by a unique and \emph{constant} value, which
is \emph{the same} for the entire range of temperatures, where the wedge
prewetting takes place. We are not aware of another study in the literature
reporting the shift of wedge prewetting due to additional confinement of the
fluid.

Using the DF approach with the arc-length continuation over the parameter $T$
(see equation (24) of Part I), we were able to calculate
$\Delta\tilde{\mu}_{\text{wpw}}\left(T\right)$ transition lines of
capillaries form figure 10 of Part I and figure \ref{Fig3} and obtain them as
sets of data in the $T$ -- $\Delta\mu$ space. However, the arc-length
continuation algorithm treats the continuation parameter as the unknown, thus
there is no control over the particular values of $T$, at which a transition
line is calculated. Thus each line was obtained at values of temperatures.
Using spline-interpolation, we have computed both
$\Delta\tilde{\mu}_{\text{wpw}}\left(T\right)$ transition lines from figure
10 of Part I and figure \ref{Fig3} at the set of 1000 points positioned
equidistantly along the intersection of the temperature ranges, where both
lines are found. The mean value of the difference between them along the
$\Delta\mu$-axis was found to be $\left(0.16\pm0.01\right)\cdot10^{-2}$.

Since the distance between both transition lines is constant (within the
margin of rounding errors), there clearly exists a special $H$-dependence of
the $\Delta\tilde{\mu}_{\text{wpw}}\left(T\right)$ transition line: at higher
values of $H$ the $\Delta\tilde{\mu}_{\text{wpw}}\left(T\right)$-line spans a
broader range of temperatures and shifts on the phase diagram \emph{entirely}
towards more negative chemical potentials. Obviously, as $H$ is increased,
the capillary corners become more isolated and behave more like infinite
wedges. At the same time, the CC behaves more like the bulk liquid -- vapour
coexistence (saturation): the $\Delta\mu_{\text{cc}}\left(T\right)$-line
approaches bulk saturation ($\Delta\mu\equiv0$), becomes more flat and also
splits into transition lines forming the loci of vapour -- capillary-liquid
and thin film -- capillary-liquid transitions (see, e.g., lines 2 and 4 in
figure \ref{Fig3}). A further increase in the value of the width $H$ of the
capped capillary will obviously lead to
$\Delta\tilde{\mu}_{\text{wpw}}\left(T\right)$-line spanning an even broader
range of temperatures. In the limit $H\to\infty$, when the capped capillary
becomes a right-angled wedge, that transition line runs \emph{tangent} to
bulk saturation (see, e.g., reference \cite{RejDietNapPRE99}, where the wedge
prewetting line, $\Delta\mu_{\text{wpw}}\left(T\right)$, is studied in
detail), and the equivalence
$\Delta\tilde{\mu}_{\text{wpw}}\left(T\right)\equiv\Delta\mu_{\text{wpw}}\left(T\right)$
takes place. We have thus shown, that there exists a \emph{unique},
\emph{constant} $H$-dependent shift of the \emph{entire}
$\Delta\mu_{\text{wpw}}\left(T\right)$-line in semi-infinite rectangular
pores. Our conclusion is based on the observed constant shift between the two
$\Delta\tilde{\mu}_{\text{wpw}}\left(T\right)$-lines of capillaries differing
only in the value of wall separation. One can possibly obtain the analytic
expression for the value of the shift, using thermodynamics methods, similar
to those outlined in, e.g., references \cite{EvMar87,Ev85}. We leave such a
study for the future.

\section{Vapour, drops and capillary-liquid slab}
\label{ThreePhaseCap}
%
%
For confined fluids the dimensions of substrate geometry, e.g., the
separation between the side walls, $H$, and the parameters of the substrate
potential, $\varepsilon_{\text{w}}$, $\sigma_{\text{w}}$, act as
thermodynamic fields controlling the phase behaviour of the fluid and have
the same status as the bulk fields $T$ and $\mu$, see, e.g., the review
\cite{HendersonBook92}. We have illustrated this concept by, first showing
how reducing the effective strength of the substrate potential (which can be
measured by, e.g., the planar wetting temperature, $T_{\text{w}}$, so that a
stronger substrate has a higher $T_{\text{w}}$) creates an additional
(metastable) phase with capillary-liquid drops in the corners (see, e.g.,
Part I, where figure 2 has a single Van der Waals loop and figure 8 which has
two Van der Waals loops). The system we investigate can serve as an
illustration of the fact that $H$ is also an independent thermodynamic field.
Increasing the wall separation of the capillary in figure 10 of Part I to get
the capillary in figure \ref{Fig3}, stabilizes the phase with corner drops,
and results in the existence of two stable first-order phase transitions:
from vapour to capillary-liquid drops (shifted wedge prewetting,
$\Delta\tilde{\mu}\left(T\right)$, transition line 6 in figure \ref{Fig3})
and then to a capillary-liquid slab (transition line 7). Also, all three
phases can coexist at a triple point at $T=T_{3}^{\text{cpd}}$.

\begin{figure}
\centering
\includegraphics{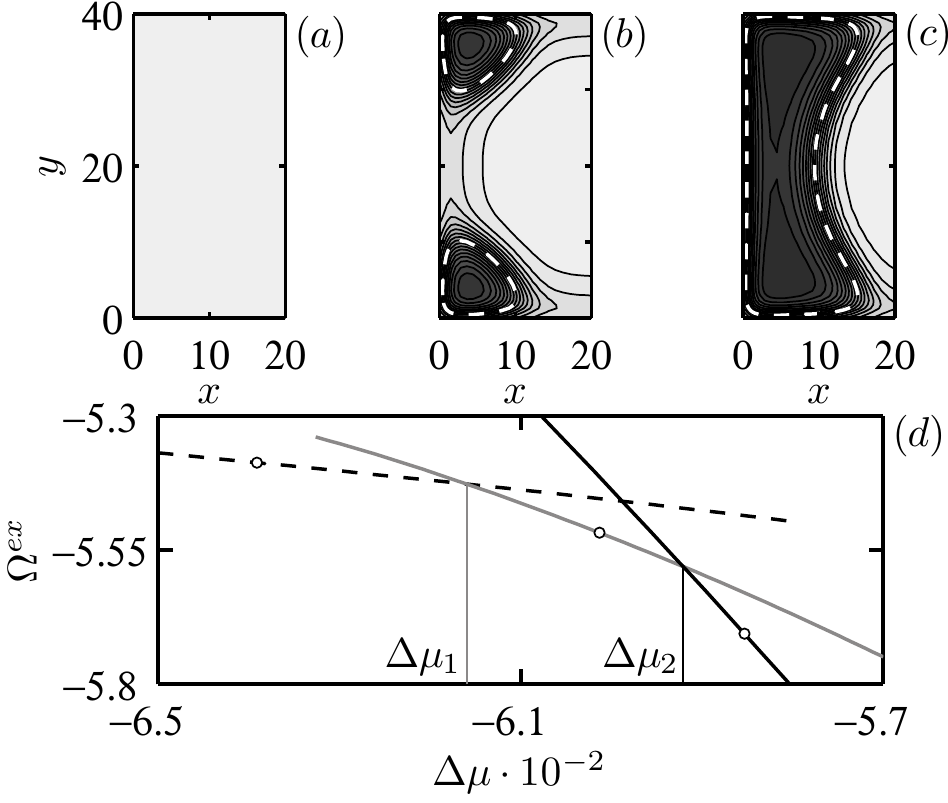}
    \caption{Isothermal thermodynamic path at $T=0.87$ ($T_3^{\text{slt}}<T<\tilde{T}_{\text{wpw}}^{\text{cr}}$),
    $\Delta\mu_{\text{cc}}\left(T\right)=-3.89\cdot10^{-2}$. Capillary parameters are given in figure \ref{Fig3}.
    As $\Delta\mu$ is increased, there are two stable consecutive first-order transitions:
    at $\Delta\mu_1\equiv\Delta\mu_{\text{cpw}}\left(T\right)=-6.15\cdot10^{-2}$ and at
    $\Delta\mu_2\equiv\Delta\tilde{\mu}_{\text{wpw}}\left(T\right)=-5.92\cdot10^{-2}$. (a) -- (c)
    Representative density profiles of fluid configurations inside each of the three stable phases.
    Reference densities: $\rho_{\text{cc}}^{\text{vap}}=0.06$, $\rho_{\text{cc}}^{\text{liq}}=0.5$.
    (d) Excess free energy isotherm (unstable branches are not shown), which possesses three concave (stable) branches,
    defining the phases: vapour (dashed line, profile at $\Delta\mu=-6.39\cdot10^{-2}$, $\Omega^{\text{ex}}=-5.39$ is shown in
    plot (a)), drop phase (solid grey line, profile at $\Delta\mu=-6\cdot10^{-2}$, $\Omega^{\text{ex}}=-5.52$ is shown in plot (b))
    and capillary-liquid slab phase (solid black line, profile at $\Delta\mu=-5.85\cdot10^{-2}$, $\Omega^{\text{ex}}=-5.71$ is shown
    in plot (c)). Values of $\left(\Delta\mu,\Omega^{\text{ex}}\right)$, corresponding to (a) -- (c) are marked on (d) by open circles.
    See also figure \ref{FigVDSis} for the adsorption isotherm.\label{FigVDSpr}}
\end{figure}

\begin{figure}
\centering
\includegraphics{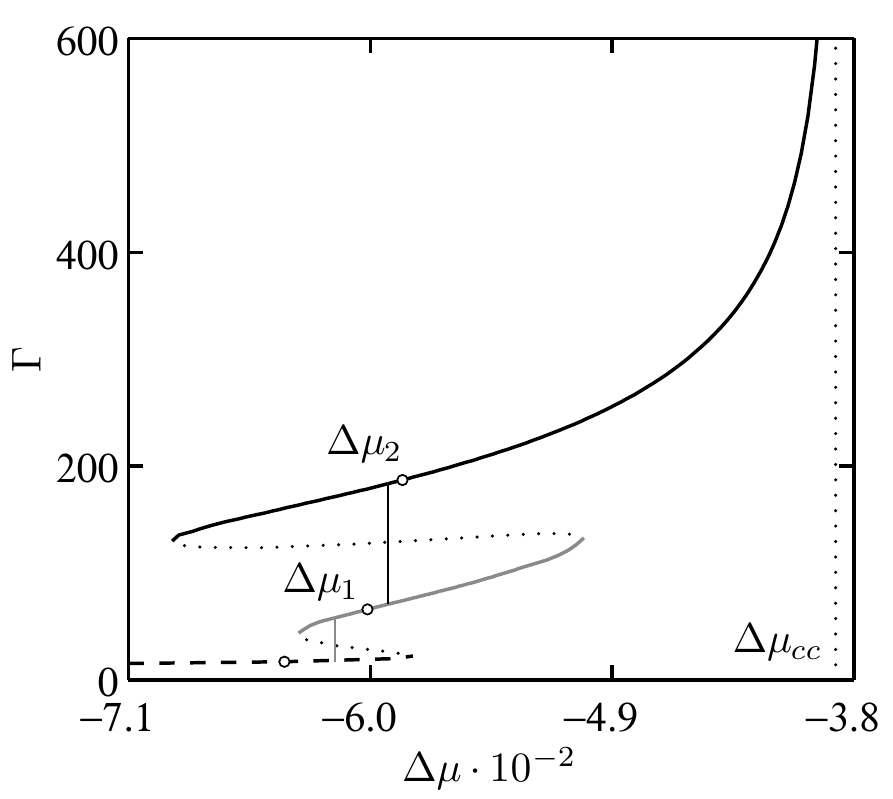}
    \caption{Adsorption isotherm, $\Gamma\left(\Delta\mu\right)$, for the thermodynamic route from figure \ref{FigVDSpr}.
    Line styles and open circles are the same as in figure \ref{FigVDSpr}(d);
    dotted branches denote thermodynamically unstable states. Dotted vertical
    line is at $\Delta\mu_{\text{cc}}\left(T\right)=-3.89\cdot10^{-2}$ and is the
    vertical asymptote for $\Gamma\left(\Delta\mu\right)$. There are two
    hysteresis loops, unlike, e.g. adsorption isotherms in figure 2 of Part I.
    The values $\Delta\mu_1$, $\Delta\mu_2$ can be obtained independently by two
    equal area constructions and correspond to jumps in adsorption: $\Delta\mu_1$
    -- from $\Gamma_1^1=16.7$ to $\Gamma_1^2=58$ (shifted wedge prewetting),
    $\Delta\mu_2$ -- from $\Gamma_2^1=70.7$ to $\Gamma_2^2=183.3$ (capillary
    prewetting).
    \label{FigVDSis}}
\end{figure}

When the shifted wedge prewetting transition is stable, there exist
isothermal thermodynamic paths exhibiting two consecutive first-order phase
transitions: shifted wedge prewetting and capillary prewetting. Consider,
e.g., an isothermal path at $T=0.87$, which vertically crosses the transition
lines 6 and 7 on the phase diagram at values of the disjoining chemical
potential $\Delta\mu_1\equiv\Delta\tilde{\mu}_{\text{wpw}}\left(T\right)$ and
$\Delta\mu_2$, passing through the areas of stability of vapour, drop and
slab phases, and ends at CC transition (line 2), at the value of disjoining
chemical potential $\Delta\mu_{\text{cc}}\left(T\right)$. The typical density
profiles inside each of the stable fluid phases are shown in figures
\ref{FigVDSpr}(a) -- \ref{FigVDSpr}(c). The isotherms of the path are
presented in figure \ref{FigVDSpr}(d) (without the unstable branches) and
figure \ref{FigVDSis}.

At low values of chemical potential the system is represented in the phase
diagram of figure \ref{Fig3} by a point below line 6, and the fluid inside
the capped capillary is in a state of vapour, figure \ref{FigVDSpr}(a).
Increasing $\Delta\mu$ isothermally corresponds to the point on the phase
diagram moving vertically up, towards line 6. Crossing this line (at
$\Delta\mu=\Delta\mu_1$) results in a discontinuous first-order phase
transition (shifted wedge prewetting), which changes the structure of the
fluid from vapour to the two capillary-liquid drops in the corners of the
capillary (figure \ref{FigVDSpr}(b)). When the point representing the system
on the phase diagram is in the region between lines 6 and 7, the
configurations with capillary-liquid drops are stable. Increasing $\Delta\mu$
towards line 7 results in the slow growth of the drops, but when the system
crosses line 7 (at $\Delta\mu=\Delta\mu_2$), there is a discontinuous
transition to a state with a capillary-liquid slab formed at the capping wall
(figure \ref{FigVDSpr}(c)). When the point of reference in the phase diagram
is in the regions between lines 7 and 2, the slab phase is stable, and
further isothermal increase of $\Delta\mu$ towards line 2 results in the slab
growing into the capillary bulk, and the system undergoes the continuous CC
at $\Delta\mu_{\text{cc}}\left(T\right)$.

Consider the excess free energy isotherm of the described thermodynamic
route, figure \ref{FigVDSis}(d). Each stable fluid phase is defined as the
concave branch of $\Omega^{\text{ex}}\left(\Delta\mu\right)$-dependence. The
branches are thus plotted with a dashed line for vapour, solid grey line for
drop and solid black line for slab phases. At any given value of $\Delta\mu$
the system should select a state with the lowest value of the free energy,
$\Omega^{\text{ex}}$. So, when the branches of
$\Omega^{\text{ex}}\left(\Delta\mu\right)$ intersect, the system follows the
minimal route along \emph{different} branches, which results in two
consecutive discontinuous first-order phase transitions in the fluid. In
figure \ref{FigVDSpr}(c) the branches of
$\Omega^{\text{ex}}\left(\Delta\mu\right)$ intersect at $\Delta\mu_1$ and
$\Delta\mu_2$, and the system follows the path along the branches of
$\Omega^{\text{ex}}\left(\Delta\mu\right)$ designated by open circles, which
denote the values of $\Delta\mu$ and $\Omega^{\text{ex}}$ corresponding to
the density profiles shown in figures \ref{FigVDSpr}(a) -- \ref{FigVDSpr}(c).
The points $\left(T,\Delta\mu_1\right)$ and $\left(T,\Delta\mu_2\right)$ thus
belong to lines 6 and 7 on the phase diagram in figure \ref{Fig3}.

We did not plot the unstable branches of the excess free energy, so as to
avoid unnecessarily complicating the figure. Such states are more relevant
for showing the continuity of an adsorption isotherm. Figure \ref{FigVDSis}
shows the dependence $\Gamma\left(\Delta\mu\right)$, where the line
specifications are the same as in figure \ref{FigVDSis}(d) and unstable
branches are plotted with a dotted line. For an isothermal thermodynamic
route the chemical potential, $\Delta\mu$, is the control parameter, while
the adsorption, $\Gamma\left(\Delta\mu\right)$, is the order parameter. A
dicontinuous phase transition thus corresponds to a jump in the value of
adsorption, as the chemical potential is being increased. Finding the
intersections of branches of the free energy is equivalent (due to the Gibbs
rule, equation (22) of Part I) to an equal area construction on the
$\Gamma\left(\Delta\mu\right)$-dependence. We have designated such
constructions by solid vertical lines at $\Delta\mu_1$ and $\Delta\mu_2$ in
figure \ref{FigVDSis}. Note that the two stable phase transitions correspond
to the two different hysteresis loops on the adsorption isotherm, unlike,
e.g., a single hysteresis loop in the case shown in figure 2(d) of Part I.
The disjoining chemical potential of CC provides the vertical asymptote for
the dependence $\Gamma\left(\Delta\mu\right)$: the point
$\left(T,\Delta\mu_{\text{cc}}\right)$ belongs to line 2 on the phase
diagram, figure \ref{Fig3}. The critical exponent for the divergence of the
order parameter at CC has been obtained in reference \cite{Parry07} using an
effective Hamiltonian formalism, and was confirmed in our previous study in
reference \cite{Yatsyshin2013} using the DF approach:
$\Gamma\sim\left(\Delta\mu-\Delta\mu_{\text{cc}}\right)^{-1/4}$.

The turning points of the $\Gamma\left(\Delta\mu\right)$-dependence -- the
spinodals -- are very important for making predictions about the behaviour of
an actual physical system, because in reality the system may remain in a
particular phase as the control parameter is changed until the spinodal of
the corresponding branch of free energy is crossed. Thus, a jump in
adsorption can happen at a value of $\Delta\mu$ anywhere in the region, where
the $\Gamma\left(\Delta\mu\right)$-dependence has more than a single value
for the given value of the chemical potential. This leads to hysteresis in
the fluid wetting behaviour: since the system can actually be found in
metastable configurations, a route from vapour to CC may be taken through
different states, than a route from CC to vapour. In section \ref{SecDyn} we
use a dynamic model to show how spinodals can pin the relaxation of the
system to its equilibrium state.

As can be seen from the phase diagram in figure \ref{Fig3}, the temperature
region where the three fluid phases can be stable is bounded from below by
the triple point, at $T\equiv T_3^{\text{cpd}}=0.865$, and from above -- by
the critical point of transition line 6, at $T\equiv
\tilde{T}_{\text{wpw}}^{\text{cr}}=0.885$. In what follows of this section we
examine in more detail these limiting cases.

\begin{figure}
\centering
    \includegraphics{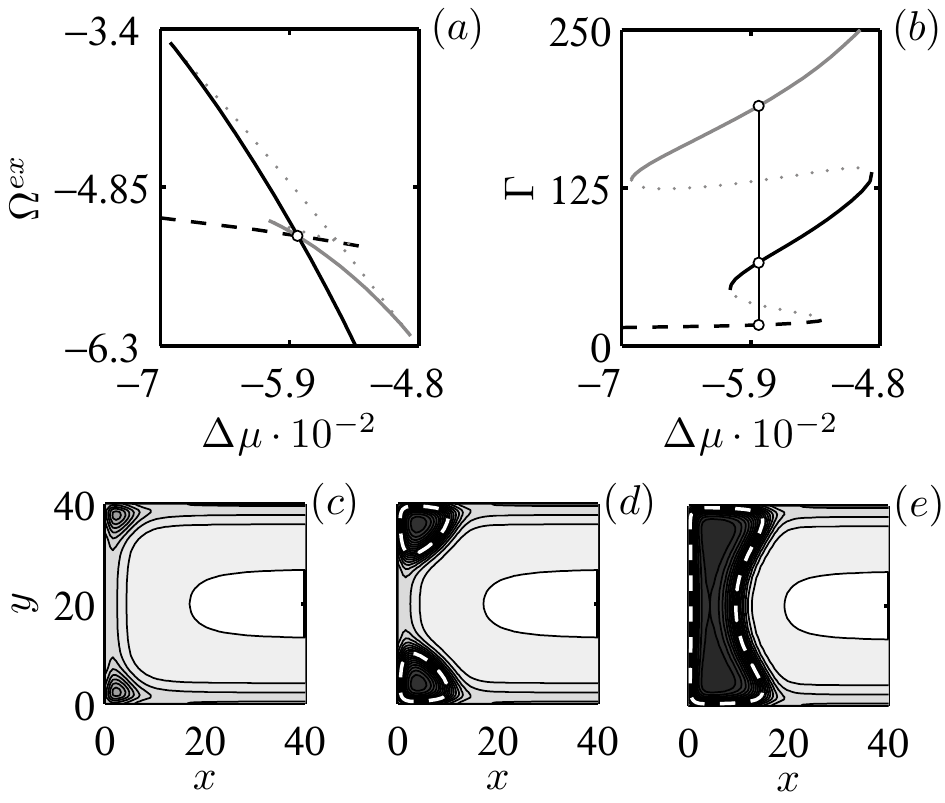}%
    \caption{Isothermal thermodynamic route at $T\equiv T_{3}^{\text{cpd}}=0.865$.
    Vapour, drop and slab phases coexist at
    $\Delta\mu_{\text{cpw}}\left(T\right)=\Delta\tilde{\mu}_{\text{wpw}}\left(T\right)=-5.83\cdot10^{-2}$.
    Capillary parameters are given in figure \ref{Fig3}. (a) and (b) Excess free energy and
    adsorption isotherms. Stable branches are defined as in figures \ref{FigVDSpr}(d) and \ref{FigVDSis}.
    Dotted line denotes the unstable branches. Open circles mark values at three-phase coexistence:
    $\Omega^{ex}=-5.29$, $\Gamma^1=16.6$, $\Gamma^2=65.8$, $\Gamma^3=189.8$. Vertical line on (b) is at the
    transition value ($\Delta\mu=-5.83\cdot10^{-2}$), which can be obtained independently by two equal area constructions.
    (c) -- (e) Density profiles of coexisting phases. Reference densities:
    $\rho_{\text{cc}}^{\text{vap}}=0.06$, $\rho_{\text{cc}}^{\text{liq}}=0.51$. \label{FigTr2D}}%
\end{figure}

Figure \ref{FigTr2D} depicts the isotherms and the density profiles of
coexisting fluid configurations corresponding to the thermodynamic route
exactly at $T\equiv T_3^{\text{cpd}}=0.865$, where vapour, drop and slab
phases coexist. The three branches of the excess free energy,
$\Omega^{\text{ex}}\left(\Delta\mu\right)$, defining the fluid phases, all
cross at a single point (see figure \ref{FigTr2D}(a)). The transition values
of the disjoining chemical potential are equal at this point:
$\Delta\mu_1=\Delta\mu_2$ (compare, e.g., figures \ref{FigTr2D}(b),
\ref{FigVDSpr}(d) and \ref{FigVDSis}). The density profiles of the three
coexisting fluid configurations are plotted in figures \ref{FigTr2D}(c) --
\ref{FigTr2D}(e). Note that the triple point exists in the capillary with
$H=40$, and does not exist in the capillary with identical substrate
potential, but with $H=30$ (see phase diagrams in figure 10 of Part I and
figure \ref{Fig3}). This happens, because a larger distance between the side
walls allows for more isolation of the capillary corners. As a consequence,
wedge wetting manifests itself here as an additional drop phase. The effect
of higher isolation of capillary corners, leading to the three-phase
coexistence can also be achieved by reducing the effective strength of the
substrate potential (by that we mean, manipulating the substrate parameters
in such a way as to lower the value of its planar wetting temperature,
$T_{\text{w}}$).

It is also quite interesting to consider the change in the triple
temperature, as the remaining thermodynamic fields, $H$ and substrate
parameters ($\varepsilon_{\text{w}}$ and $\sigma_{\text{w}}$), are varied.
Obviously, increasing $H$, or reducing the strength of the substrate will
lead to higher isolation of corners, and the triple point will remain. But
reducing $H$ or increasing the substrate strength will lead to the appearance
of the tri-critical point in the system, where the three-phase coexistence
terminates. The apparent existence of the tri-critical point is again proven
by the two examples, in figure 10 of Part I and in figure \ref{Fig3}, where
the first one exhibits a triple-point, and the second one does not. We leave
an investigation of the triple-line for future studies.

Consider now the higher-temperature boundary of the region in the phase
diagram, where vapour, drop and slab phases can be stable at the same $T$
(but different values of $\Delta\mu$). It is set by the critical point of
line 6, at temperature $\tilde{T}_{\text{wpw}}^{\text{cr}}=0.885$. We
contrast the behaviour of the fluid near that critical point with the
behaviour near the low-temperature end of line 6, at the temperature
$\tilde{T}_0=0.828$, by considering two isothermal routes. We also show how a
mean-field DF approach is capable of capturing a signature of the approach to
criticality.

\begin{figure}
\centering
    \includegraphics{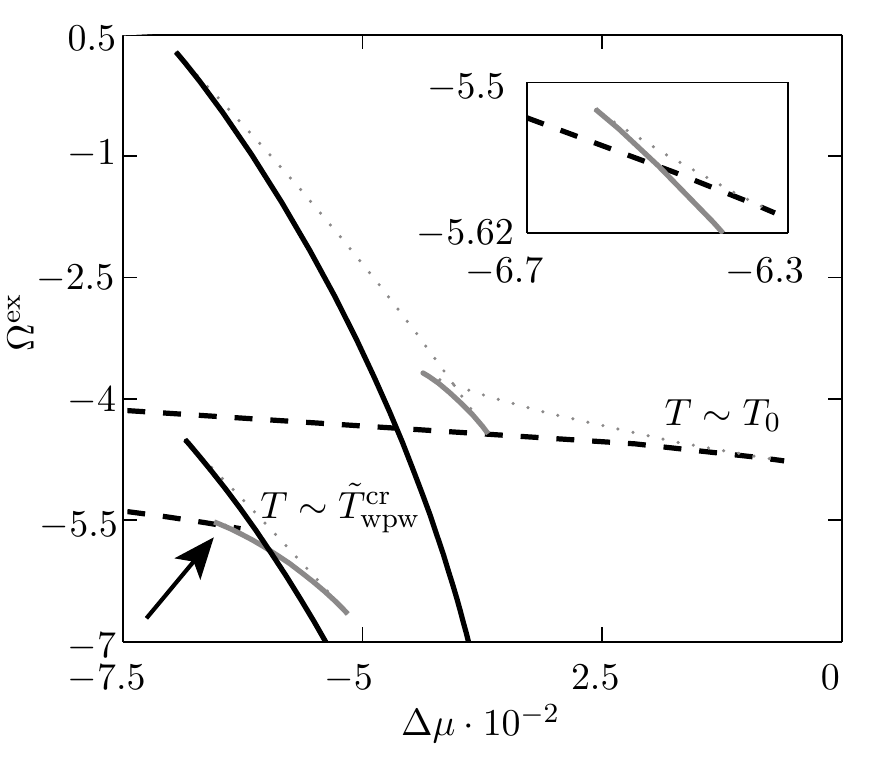}%
    \caption{Criticality of the shifted wedge prewetting. The figure shows two excess free energy isotherms at
    values of $T$ near the ends of
    $\Delta\tilde{\mu}_{\text{wpw}}\left(T\right)$-line (line 6) from figure
    \ref{Fig3}, each possessing two Van der Waals loops: of shifted wedge
    prewetting and of capillary prewetting. Each isotherm has three concave
    branches (connected by non-concave branches, plotted with dotted lines),
    defining the phases: vapour (dashed line), corner drops (solid grey) and
    capillary-liquid slab (solid black). Arrow points at the lower isotherm with
    the near-critical shifted wedge prewetting, at
    $T_1=0.875\lesssim\tilde{T}_{\text{wpw}}^{\text{cr}}=0.885$ (higher-$T$ end
    of line 6 in figure \ref{Fig3}), with transition at point
    $\left(\Delta\tilde{\mu}_{\text{wpw}}\left(T_1\right),\Omega^{\text{ex}}\right)=\left(-6.5\cdot10^{-2},\right)$
    (capillary prewetting is at point
    $\left(\Delta\mu_{\text{cpw}}\left(T_1\right),\Omega^{\text{ex}}\right)=\left(6\cdot10^{-2},-5.87\right)$).
    Inset zooms on the shifted wedge prewetting transition at
    $\Delta\tilde{\mu}_{\text{wpw}}\left(T_1\right)$, which disappears in the
    limit $T\to\tilde{T}_{\text{wpw}}^{\text{cr}}$, as the branches of excess
    free energy defining vapour and drop phases align and form a single branch
    defining vapour. The upper isotherms is at $T_2=0.83\gtrsim
    \tilde{T}_0=0.828$ (lower-$T$ end of line 6 in figure \ref{Fig3}), with
    transition at point
    $\left(\Delta\tilde{\mu}_{\text{wpw}}\left(T_2\right),\Omega^{\text{ex}}\right)=\left(-4.7\cdot10^{-2},-4.36\right)$
    (capillary prewetting is at point
    $\left(\Delta\mu_{\text{cpw}}\left(T_2\right),\Omega^{\text{ex}}\right)=\left(-4.7\cdot10^{-2},-4.36\right)$).
    The shifted wedge prewetting transition disappears at $T=\tilde{T}_0$,
    because the branch of excess free energy defining drop phase losses
    intersection with the branch defining vapour. Thus, the temperature
    $\tilde{T}_0$ is not critical, whereas the temperature
    $\tilde{T}_{\text{wpw}}^{\text{cr}}$ is critical.
    \label{FigCritical}}%
\end{figure}

Figure \ref{FigCritical} shows two free-energy isotherms: at
$T=0.875\lesssim\tilde{T}_{\text{wpw}}^{\text{cr}}=0.885$ (the lower
isotherm, designated by an arrow, which also points to the area zoomed at on
the inset) and at $T=0.83\gtrsim \tilde{T}_0=0.828$ (the upper isotherm).
Consider, first, the lower isotherm. As can be seen from the inset, the
vapour -- drop transition is present and is stable, but it is also clear
(compare to, e.g. figure \ref{FigVDSpr}(d), which shows an isotherm at a
relatively lower temperature, $T=0.87$), that the branches of excess free
energy, defining the vapour and drop phases, tend to align and form a single
branch, as $T\to\tilde{T}_{\text{wpw}}^{\text{cr}}$. In the limit
$T=\tilde{T}_{\text{wpw}}^{\text{cr}}$ the distinction between vapour and
drop phases will be lost, as branches of excess free energy will join
together and form one continuous branch of
$\Omega^{\text{ex}}\left(\Delta\mu\right)$, extending from $\mu=-\infty$.
During an isothermal increase of the chemical potential at
$T>\tilde{T}_{\text{wpw}}^{\text{cr}}$ the structure of the fluid changes
continuously from vapour to a configuration with drops in the corners.

The structure of the upper isotherm in figure \ref{FigCritical} is
principally different. As is clear from the figure, the branch of excess free
energy, which defines the drop phase (solid grey line), simply nearly loses
the intersection point with the branch defining the vapour phase (dashed
line). Generally, when lowering the temperature from
$\tilde{T}_{\text{wpw}}^{\text{cr}}$ and computing consecutive isothermal
thermodynamic routes, we observe that the branch of excess free energy, which
defines the drop phase (solid grey line), moves up with respect to the one
defining vapour (dashed line).  The intersection is thus lost in the limit
$T\equiv\tilde{T}_0=0.828$ (compare with isotherms computed at higher $T$:
see lower isotherm in figure \ref{FigCritical} ($T=0.875$), as well as
isotherms in figures \ref{FigVDSpr}(d) ($T=0.87$) and \ref{FigTr2D}(a)
($T=0.865$)). A concave branch of excess free energy, defining the drop phase
can still be found on the isotherms at $T\lesssim\tilde{T}_0$, but it has no
intersection with any other branches, thus even a metastable phase transition
is lost. The vapour -- drop transition line
($\Delta\tilde{\mu}_{\text{wpw}}\left(T\right)$, line 6 in figure \ref{Fig3},
including its metastable part) can never cross the CC line, because the
density profiles $\rho^{\text{cpd}}\left(x,y\right)$ with corner drops, can
be solutions to the Euler-Lagrange equation (see equation (11) of Part I)
only when the capillary bulk is filled with vapour, and such configurations
are unstable above the CC line, where the bulk is filled with
capillary-liquid.

The DF method, which we are using, is classical in the sense that it assumes
the molecular positions to be fixed at their ensemble averages. The main
limitation of such an approach is that the thermal fluctuations of particles
around their equilibrium positions are completely neglected. As a result, the
criticality, where such fluctuations play a dominant role, cannot be
accounted for within a classical mean-field approach. However, as we have
demonstrated, the analysis of Van der Waals loops allows us to capture a
signature of the approach to criticality, and that precisely is how we
located the critical points on all presented phase diagrams (see, e.g., open
circles in figure \ref{Fig3}). Once a critical behaviour is known to exist,
it can be studied with more appropriate methods, such as effective
Hamiltonians and renormalisation group theory, see, e.g.,
\cite{TasinkevychDietrichEurPhysJE07, MilchevPRL03, LikosPhysRep01,
RasconParrtJChemPhys99, DarbellayYeomans92}. We leave such studies for the
critical points at $T=T_{\text{cpw}}^{\text{cr}}$ and
$T=\tilde{T}_{\text{wpw}}^{\text{cr}}$ (e.g., end points of lines 6 and 7 in
figure \ref{Fig3}) for future research.

\section{Three-phase coexistence in the slit pore}
All the examples up to this point were restricted to cases where at a given
temperature the capillary bulk (associated slit pore) can have at most two
stable phases: vapour or capillary-liquid. We have shown that a transition
between these phases changes dramatically, when the slit pore is capped by an
additional wall and a capillary wetting temperature appears as a result.
However, a slit pore can allow for the existence of a third phase, a thin
film, which is a remnant of the prewetting on a single wall, see, e.g.,
\cite{Dietrich}. In the next section we show, how such a complex behaviour of
capillary bulk leads to a continuous prewetting transition in the capped
capillary. Here we provide some background and briefly describe three-phase
coexistence in the slit pore, using as an example the 1D slit pore whose full
phase diagram is shown in figure \ref{Fig3} (solid and dashed grey lines).
First, we briefly summarise the phenomenology of adsroption on a single wall
and then describe the changes due to adding another wall to form a slit pore.
Finally, we discuss the three-phase coexistence inside the capillary bulk and
provide several representative examples associated with various isothermal
thermodynamic routes near $T_3^{\text{slt}}$, figure \ref{Fig3}.

Although wetting on a single wall and a slit pore is well understood, and has
attracted considerable interest in the literature from the late 1980's until
the mid 1990's (see, e.g., reviews in
\cite{EvansParry90,Yatsyshin2012,MonsonMicroMesoMat12}), we believe it is
appropriate and within the scope of the present study to provide a treatment
of single-wall and slit-like substrates within our DF approach. Such
treatment reveals the connections between the wetting scenarios of the 1D
slit pore and 2D capped capillary, as well as providing the necessary
background for studying the effect of capping a capillary, whose bulk can
have multiple stable phases. In the existing literature numerical studies
most relevant to the present section are, e.g., reference
\cite{BrunoMarconiEvansPhysicaA1987}, where the three-phase coexistence in a
slit pore is studied using a mean-field microscopic lattice-gas model of the
fluid. The three phase coexistence is mentioned in references
\cite{Evans86,EvansParry90}. The treatment of asymmetric slit-like
substrates, using effective substrate potentials, as well as a non-local DF
model, can be found in the recent work by Stewart and
Evans~\cite{StewartEvansPhysRevE12}.


In the case of a single planar wall immersed in vapour the bulk liquid --
vapour transition has the option of becoming continuous (second-order), i.e.,
it can happen from the substrate surface. For example, in an isothermal
approach to bulk saturation ($\Delta\mu\equiv0$), at $T$ higher than the
planar wetting temperature, $T_{\text{w}}$, the space above the wall is
filled with liquid continuously, as the chemical potential approaches
saturation. The liquid -- vapour transition is then preceded by prewetting, a
first-order surface phase transition between the vapour and a thin
(microscopic) film adsorbed on the substrate wall. Transition line 1 in
figure \ref{Fig3} is the locus of prewetting transitions on the planar
substrate -- $\Delta\mu_{\text{pw}}\left(T\right)$. At values of $T$ below
$T_{\text{w}}$ the liquid -- vapour transition follows its bulk scenario
(first-order): wall stays non-wet (non-zero contact angle), until the
chemical potential reaches saturation, and the whole system abruptly
transforms to a liquid state.

Confining a fluid which is in its vapour phase in a slit pore adds wall
separation and parameters of the substrate potential to the set of bulk
thermodynamic fields (temperature and chemical potential). The phenomenology
of phase transition in the confined fluid maps to that in the bulk: the
transition between vapour and capillary-liquid is of first order. But the
denser coexisting phase (capillary-liquid) has lower density (far from the
surfaces of both walls, where $\rho^{\text{slt}}\left(y\right)$ is nearly
constant) than the density of the bulk liquid, coexisting with vapour at the
same temperature. Moreover, the transition in the slit pore is essentially
the shifted bulk liquid -- vapour transition. This phenomenon is otherwise
referred to as the Kelvin shift and was discussed earlier, in the
Introduction. In the simplest macroscopic description, where all non-local
fluid-substrate intermolecular effects are contained inside surface tensions,
the vapour -- capillary-liquid transition is given by equation (1) of Part I.
The CC lines, which we present in this work (e.g., grey curves in figures 4,
9, 10 of Part I and in figure \ref{Fig3}), describe the same phenomenon, but
include all the fluid-fluid and fluid-substrate molecular interactions
explicitly, in the model of the fluid free energy (equation (2) of Part I),
which is equivalent to approximating the grand partition function.
\begin{figure}
\centering
    \includegraphics{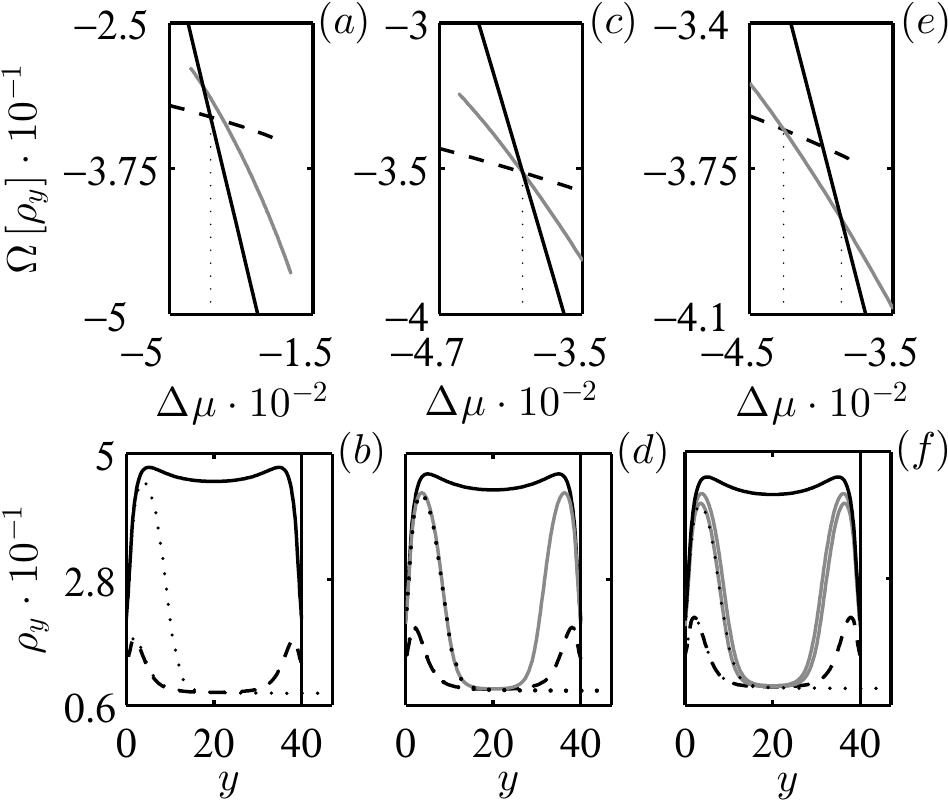}%
    \caption{Stable transitions during adsorption on the slit pore near its triple temperature, $T_3^{\text{slt}}=0.92$
    (see full phase diagram of the slit pore in figure \ref{Fig3} plotted with grey lines and the planar prewetting line
    plotted with dotted grey line). (a) and (b) $T=0.910<T_3^{\text{slt}}$, coexisting phases: vapour -- capillary-liquid,
    at $\Delta\mu=-4\cdot10^{-2}$, $\Omega\left[\rho^{\text{slt}}\right]=-3.3\cdot10^{-1}$.
    (c) and (d) $T=0.920\equiv T_3^{\text{slt}}$, coexisting phases: vapour -- thin film -- capillary-liquid,
    at $\Delta\mu=-4\cdot10^{-2}$, $\Omega\left[\rho^{\text{slt}}\right]=-3.5\cdot10^{-1}$. (e) and
    (f) $T=0.927>T_3^{\text{slt}}$, coexisting phases: vapour -- thin film,
    at $\Delta\mu=-4.3\cdot10^{-2}$, $\Omega\left[\rho^{\text{slt}}\right]=-3.7\cdot10^{-1}$, and
    thin film -- capillary-liquid, at $\Delta\mu=-9\cdot10{-2}$, $\Omega\left[\rho^{\text{slt}}\right]=-3.9\cdot10^{-1}$.
    In (a), (c), (e) unstable branches of free energy are not plotted and
    dotted vertical lines mark values of $\Delta\mu$, where stable transitions take place.
    Branches of free energy isotherms are plotted with different line styles to distinguish between
    vapour (dashed black), capillary-liquid (solid black) and thin film (solid grey) phases.
    In (b), (d), (f) the density profiles of coexisting phases are plotted with the same line styles as the
    branches of free energy defining the phases. Black vertical lines mark the position of the second side wall of
    the slit pore, the first side wall being at $y=0$. The dotted density profiles are those of planar prewetting
    (at the same values of $T$), which takes place at
    (b) $\Delta\mu=-3.5\cdot10^{-2}$; (d) $\Delta\mu=-3.8\cdot10^{-2}$; (f) $\Delta\mu=-4.1\cdot10^{-2}$. \label{FigTrPr1}}
\end{figure}

A slit pore of any width, $H$, allows for the coexistence between vapour and
capillary-liquid phases. Increasing $H$ leads to higher isolation of the
walls and can result in the appearance of stable thin film adsorbed on both
walls, coexisting with vapour, as it happens in the case of a single wall. An
example of two slit pores, differing only in the values of wall separation,
is provided by (see grey lines) figure 10 of Part I ($H=30$: no thin film --
capillary-liquid transition) and figure \ref{Fig3} ($H=40$: stable thin film
-- capillary-liquid transition line appears,
$\Delta\tilde{\mu}_{\text{pw}}\left(T\right)$, line 3). In the latter case,
the three two-phase coexistence lines (2, 3 and 4) cross at a \emph{single
point}, at the slit triple temperature, $T_3^{\text{slt}}=0.92$, where all
three phases (vapour, thin film and capillary-liquid) are stable and coexist.

The transition line where vapour coexists with a thin film on a single planar
wall immersed in vapour (the planar prewetting line,
$\Delta\mu_{\text{pw}}\left(T\right)$, line 1 in figure \ref{Fig3}), is
shifted in the slit  pore \emph{entirely} by a constant value and forms the
transition line $\Delta\tilde{\mu}_{\text{pw}}\left(T\right)$ (line 3). An
analytical relation between the shift of prewetting and the width of the
capillary, $H$, can be obtained using standard field-theoretical
thermodynamic methods, see, e.g., references \cite{EvMar87,Ev85}. In the case
presented on figure \ref{Fig3} the transition lines of planar prewetting,
$\Delta\mu_{\text{pw}}\left(T\right)$ (line 1, figure \ref{Fig3}), and of
shifted planar prewetting in the slit pore,
$\Delta\tilde{\mu}_{\text{pw}}\left(T\right)$ (line 3), were found by
arc-length continuation of solutions to the equation (24) of Part I over the
parameter $T$, and are thus given at different values of temperature. Using
spline interpolation, we have computed both transition lines at the set of
1000 equidistant points along the $T$-axis, between both ends of
$\Delta\tilde{\mu}_{\text{pw}}\left(T\right)$-line (line 3), i.e. $T=0.848$
and $T=0.945\equiv\tilde{T}_{\text{pw}}^{\text{cr}}$. The calculated mean
value for the shift of planar prewetting transition due to the presence of
the second wall, forming the slit pore, along $\Delta\mu$-axis is found to be
$\left(1.7\pm0.1\right)\cdot10^{-3}$.

Due to the presence of the triple point, the transition to capillary-liquid
takes place along lines 2 and 4 in figure \ref{Fig3}, which together form the
$\Delta\mu_{\text{cc}}\left(T\right)$-line, where capillary-liquid coexists
with vapour (at $T<T_3^{\text{slt}}=0.92$, line 2) and thin film (at
$T>T_3^{\text{slt}}=0.92$, line 4) phases. Figure \ref{FigTrPr1} shows
representative free energy isotherms, $\Omega\left(\Delta\mu\right)$,
(unstable branches are not plotted) and coexisting density profiles (of only
stable transitions) at temperatures below, at and above $T_3^{\text{slt}}$.
The dashed line, solid black line and grey line in the plots of the isotherms
and density profiles demarcate the vapour phase, the capillary-liquid phase,
and the thin film phase, respectively. The density profiles plotted with a
dotted line show the fluid configurations coexisting during planar prewetting
on a single planar wall (the locus of phase transitions is
$\Delta\mu_{\text{pw}}\left(T\right)$-line, line 1 in figure \ref{Fig3}) at
the same value of temperature.

When $T<T_3^{\text{slt}}$ (figures \ref{FigTrPr1}(a) and \ref{FigTrPr1}(b))
there is a single stable phase transition -- between configurations of vapour
and capillary-liquid. Figure \ref{FigTrPr1}(a) shows the complete branch of
the free energy defining the thin film phase (solid grey line). It is bounded
by its two turning points (spinodals). The branch defining the vapour phase
(dashed line) extends from $\mu=-\infty$ up to its spinodal, whereas the
capillary-liquid branch is bounded at higher values of $\Delta\mu$ by the
bulk liquid -- vapour coexistence ($\Delta\mu\equiv0$). Increasing the
temperature results in the thin film branch moving down relatively to the
intersection of the vapour and capillary-liquid branches, as is clear from
figures \ref{FigTrPr1}(c) and \ref{FigTrPr1}(e).

At $T=T_3^{\text{slt}}$ (figures \ref{FigTrPr1}(c) and \ref{FigTrPr1}(d)) the
three branches of the free energy intersect at one point, at the value of
disjoining chemical potential
$\Delta\mu_3\equiv\Delta\tilde{\mu}_{\text{pw}}\left(T_3^{\text{slt}}\right)=\Delta\mu_{\text{cc}}\left(T_3^{\text{slt}}\right)$.
The triple point, $\left(\Delta\mu_3,T_3^{\text{slt}}\right)$, depends on the
substrate parameters, $\varepsilon_{\text{w}}$, $\sigma_{\text{w}}$ and the
capillary width, $H$. The dependence (triple-line) can be obtained
numerically by using arc-length continuation applied to equation
\eref{3point}, with the parameter of interest. Finally, note that the density
profiles of vapour and thin film phases in the slit pore (dashed and solid
grey lines) are almost indistinguishable from those on a single planar wall
(dotted lines).

At $T>T_3^{\text{slt}}$ (figures \ref{FigTrPr1}(e) and \ref{FigTrPr1}(f)) we
find two stable transitions: vapour -- thin film (at a lower value of
$\Delta\mu$), and thin film -- capillary-liquid (at a higher value of
$\Delta\mu$). The density profiles of vapour in the slit pore and on the
single wall are still very close. As for thin film profiles, the one on a
single wall is closer to the one in the slit pore, which coexists with
vapour, than the one coexisting with capillary-liquid.

A further increase in $T$ leads to the thin-film -- vapour criticality at the
value of shifted critical prewetting temperature,
$T=\tilde{T}_{\text{pw}}^{\text{cr}}$. The density profiles become
indistinguishable, with the corresponding phase branches of
$\Omega\left(\Delta\mu\right)$ aligning. At a fixed
$T>\tilde{T}_{\text{pw}}^{\text{cr}}$ the growth of the film adsorbed on each
wall of the slit pore is continuous with $\Delta\mu$, until the first-order
CC is reached and the entire pore is filled with capillary-liquid
discontinuously.

\section{Continuous prewetting}

In the previous section we have demonstrated (in agreement with existing
theory, e.g., references \cite{EvansParry90,Evans86}), that the prewetting of
a single planar wall immersed in vapour, and the shifted prewetting of a slit
pore immersed in vapour, are essentially equivalent phenomena. The transition
line, which forms the locus of vapour and film coexistence in a slit pore,
$\Delta\tilde{\mu}_{\text{pw}}\left(T\right)$, is shifted \emph{entirely}
from the prewetting line of the planar wall,
$\Delta\mu_{\text{pw}}\left(T\right)$, by a well defined constant value
related to the substrate parameters and pore width (see lines 1 and 3 in
figure \ref{FigTr2D} and theory in, e.g., reference \cite{EvMar87}). Also,
the calculated structure of the film phases formed in a slit and on a planar
wall is nearly indistinguishable (see, e.g., figure \ref{FigTrPr1}). In this
section we exploit this equivalence to demonstrate a case of continuous
prewetting transition and highlight some important potential applications of
this newly discovered phenomenon.

We focus on a capped capillary, whose bulk (the associated slit pore) allows
for the coexistence between vapour and thin films adsorbed on the side walls.
First, we show that the shifted prewetting transition becomes a continuous
phenomenon in such systems, manifested by capillary-liquid fingers unbinding
into the capillary bulk, as the chemical potential (pressure) is increased
towards the value at shifted prewetting. Second, we show that an isothermal
approach to CC goes through two consecutive \emph{stable} continuous phase
transitions. We are not aware of a previous study on continuous prewetting in
a confined fluid. Below we also briefly discuss some obvious similarities to
wetting of micro-steps (see, e.g., reference \cite{SaamJLowTempPhys09}). It
is to be noted, however, that the fluids under consideration are confined and
undergo saturation below the bulk liquid -- vapour transition (due to Kelvin
shift, e.g., equation (1) of Part I), unlike a step-shaped substrate immersed
in vapour. Also, we expect the continuous prewetting in the capillary-like
geometries to be much more accessible experimentally, than step wetting,
because it requires much fewer conditions to be satisfied by the substrate in
order to be observed. We elaborate on that at the end of the section.

Consider an isothermal thermodynamic route at $T=T_1$ in the interval
$T_3^{\text{slt}}<T_1<\tilde{T}_{\text{pw}}^{\text{cr}}$, where the pressure
in the reservoir (or, equivalently, the disjoining chemical potential,
$\Delta\mu$) is slowly increased from a low value, at which the fluid inside
the capped capillary is in the vapour phase. In the phase diagram this
corresponds to a path crossing vertically the transition lines
$\Delta\tilde{\mu}_{\text{pw}}\left(T\right)$ and
$\Delta\mu_{\text{cc}}\left(T\right)$ (lines 3 and 4 in figure \ref{Fig3}).
As we have seen in the previous section, an isothermal crossing of
$\Delta\tilde{\mu}_{\text{pw}}\left(T\right)$-line corresponds to the phase
transition in the capillary bulk, where vapour and thin film phases coexist
(shifted prewetting transition, see, e.g., figures \ref{FigTrPr1}(b),
\ref{FigTrPr1}(e)). One might intuitively expect that due to the presence of
the capping wall this transition may happen continuously, analogously to the
continuous CC (provided that the capping wall is sufficiently attractive). We
indeed find it to be the case.

\begin{figure}
\centering
\includegraphics{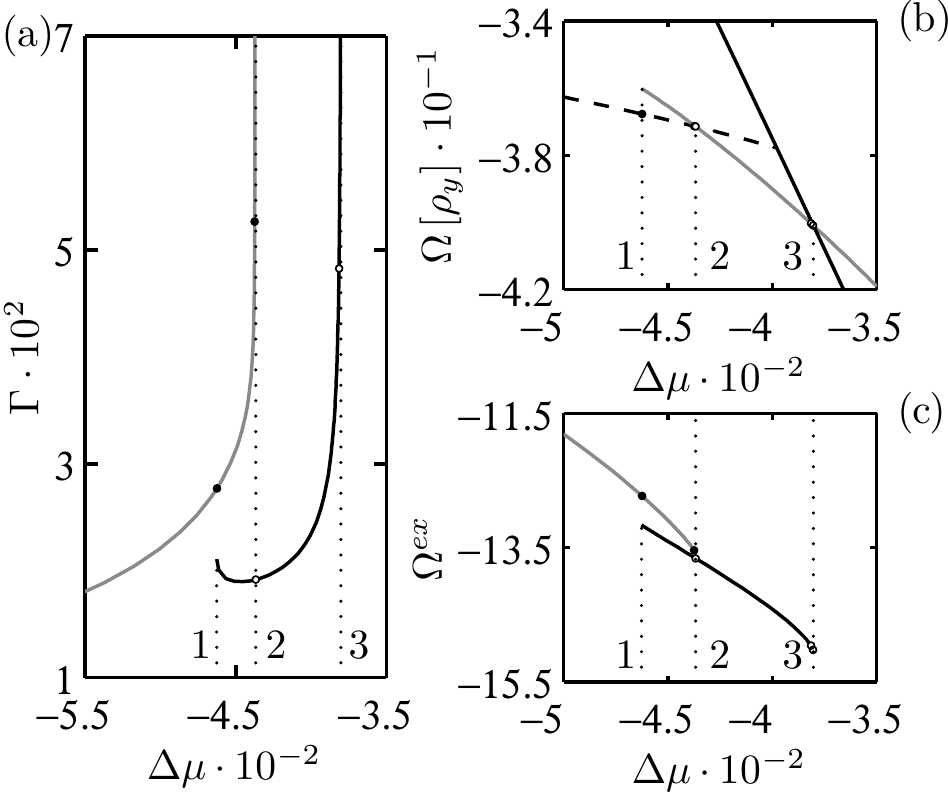}
    \caption{Isothermal thermodynamic route at $T=0.93$, across transition lines 3
    ($\Delta\tilde{\mu}_{\text{wpw}}\left(T\right)$, shifted prewetting) and 4
    ($\Delta\tilde{\mu}_{\text{cc}}\left(T\right)$, CC), see figure \ref{Fig3}.
    There are two consecutive continuous transitions: continuous prewetting, at
    $\Delta\mu_2\equiv\Delta\tilde{\mu}_{\text{wpw}}\left(T\right)=-4.37\cdot10^{-2}$,
    followed by continuous CC, at
    $\Delta\mu_3\equiv\Delta\tilde{\mu}_{\text{wpw}}\left(T\right)=-3.8\cdot10^{-2}$.
    (a) Adsorption isotherm of the capped capillary. It consists of two diverging
    branches: grey line and black line mark continuous prewetting and
    continuous CC, respectively. (b) Free energy isotherm of associated slit pore (capillary
    bulk), where (see figure \ref{FigTrPr1}) black dashed line denotes vapour,
    solid grey -- thin film, solid black -- capillary-liquid phases of the slit
    pore. (c) Excess free energy isotherm of the capped capillary; branches
    defined as in (a). Continuous prewetting happens at
    $\Omega^{\text{ex}}\approx-13.55$, continuous CC happens at
    $\Omega^{\text{ex}}\approx-15.03$. In (a) -- (c) the vertical dotted lines
    are at values $\Delta\mu_1=-4.6\cdot10^{-2}$ (spinodal of thin film branch in
    the slit pore), $\Delta\mu_2$ and $\Delta\mu_3$. Representative density
    profiles are plotted in figure \ref{FigSecTrPr}, with corresponding points
    marked on the isotherms by black circles for profiles (b) -- (c) and open
    circles for profiles (e) -- (f). \label{FigSecTrIs}}
\end{figure}

\begin{figure}
\centering
\includegraphics{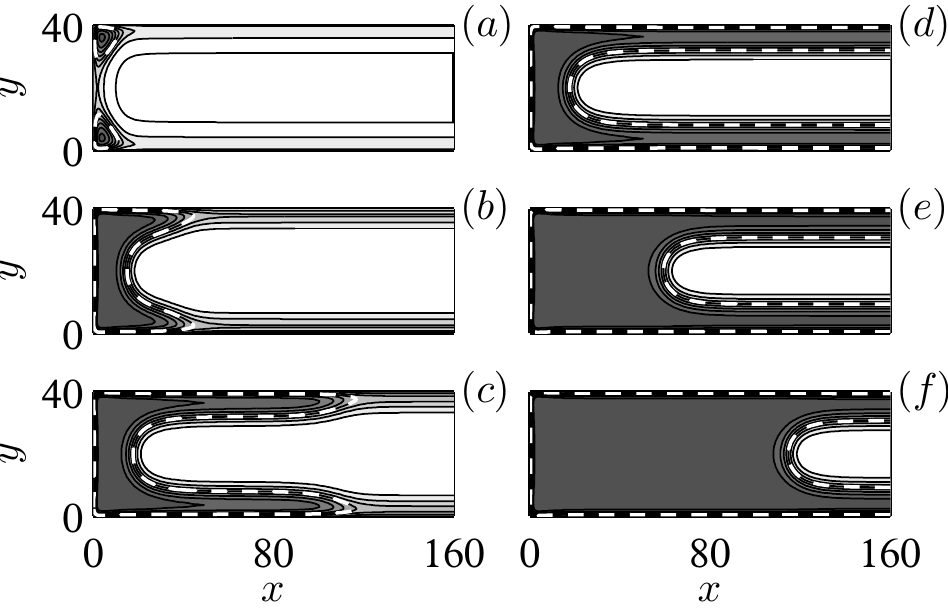}
    \caption{Representative density profiles for an isothermal approach to CC at $T=0.93$,
    which has two consecutive continuous transitions, see figure \ref{FigSecTrIs}.
    Reference densities: $\rho_{\text{cc}}^{\text{vap}}=0.09$, $\rho_{\text{cc}}^{\text{liq}}=0.41$. (a) -- (c)
    Density profiles of fluid states from the branch of the isotherm corresponding to continuous prewetting
    (figure \ref{FigSecTrIs}, grey line). The values of
    $\Delta\mu$: $-4.4\cdot10^{-2}$, $-3.82\cdot10^{-2}$, $-3.80\cdot10^{-2}$;
    values of $\Gamma$: 55.4, 276.8, 527.9; values of $\Omega^{\text{ex}}$: -7.78, -12.73, -13.54; .
    (d) -- (f) Density profiles of continuous CC, which follows continuous prewetting
    (figure \ref{FigSecTrIs}, black line). The values of
    $\Delta\mu$: $-8.3\cdot10^{-2}$, $-4.6\cdot10^{-2}$, $-4.37\cdot10^{-2}$;
    values of $\Gamma$: 193.6, 483.6, 898.1; values of $\Omega^{\text{ex}}$: -13.65, -14.96, -15.02.
    The values of adsorption and excess free energy are marked in figures \ref{FigSecTrIs}(a),
    \ref{FigSecTrIs}(c) by filled black circles for profiles (b), (c) and by open circles for profiles (d), (e). \label{FigSecTrPr}}
\end{figure}

Figure \ref{FigSecTrIs} shows a representative set of adsorption and excess
free energy isotherms, corresponding to the thermodynamic route at the
temperature $T_1=0.93>T_3^{\text{slt}}$. The region in $\Delta\mu$ is chosen
in the vicinity of transitions in the capillary bulk. Figure
\ref{FigSecTrIs}(a) shows the adsorption isotherm which consists of two
unconnected branches: the grey branch spans the interval of values of
$\Delta\mu:-\infty\leq\Delta\mu\leq\Delta\mu_2$, and the black branch is
limited to the interval $\Delta\mu:\Delta\mu_1\leq\Delta\mu\leq\Delta\mu_3$.
Figure \ref{FigSecTrIs}(b) shows the free energy isotherm of the associated
slit pore (capillary bulk), corresponding to the same thermodynamic route;
unstable branches are not shown. In figure \ref{FigSecTrIs}(b) the dashed,
solid grey and solid black like denotes the branch of vapour, thin film and
capillary-liquid, respectively. Figure \ref{FigSecTrIs}(c) shows the excess
free energy isotherm of the capped capillary, which consists of two
unconnected branches. The numbered dotted vertical lines in each figure are
drawn at the values of $\Delta\mu$, which denote: 1 -- $\Delta\mu_1$, the
value of low-pressure spinodal of the thin film phase inside the slit (the
point $\left(T_1,\Delta\mu_1\right)$ is below, but near line 3 in the phase
diagram), 2 --
$\Delta\mu_2\equiv\Delta\tilde{\mu}_{\text{pw}}\left(T_1\right)$, the value
of vapour -- thin film coexistence in the capillary bulk (shifted prewetting
transition: the point $\left(T_1,\Delta\mu_2\right)$ belongs to line 3 in the
phase diagram), 3 --
$\Delta\mu_3\equiv\Delta\mu_{\text{cc}}\left(T_1\right)$, the value of thin
film -- capillary-liquid coexistence in the capillary bulk (CC: the point
$\left(T_1,\Delta\mu_3\right)$ belongs to line 4 in the phase diagram). The
capillary bulk transitions from vapour to thin film and from thin film to
capillary-liquid are both stable. In the capped capillary $\Delta\mu_2$ and
$\Delta\mu_3$ provide vertical asymptotes for the diverging adsorption,
$\Gamma\left(\Delta\mu\right)$, plotted in figure \ref{FigSecTrIs}(a).
Several representative density profiles are shown in figure \ref{FigSecTrPr},
where plots (a) -- (c) correspond to $\Delta\mu\to\Delta\mu_2$ and plots (d)
-- (f) correspond to $\Delta\mu\to\Delta\mu_3$. The data values on the
isotherms, corresponding to the presented density profiles are marked by
filled (for profiles (b), (c)) and open (for profiles (d), (e)) circles in
figures \ref{FigSecTrIs}(a) -- \ref{FigSecTrIs}(c).

Let us describe the way in which the two consecutive continuous phase
transitions take place in the capped capillary, as the chemical potential of
the containing reservoir is increased from a large negative value
$\Delta\mu\sim-\infty$ (where the capped capillary is filled entirely with
vapour) to the value $\Delta\mu=\Delta\mu_3$ (where the capped capillary is
filled entirely with capillary-liquid), at the constant temperature
$T_1=0.93$.

As $\Delta\mu$ is increased from a large negative value so that the point
representing the system in the phase diagram, approaches and crosses the
$\Delta\mu_{\text{cpw}}\left(T\right)$-line (line 7 in figure \ref{Fig3}),
the discontinuous (first-order) transition from the drop phase to
capillary-liquid slab phase takes place at the value
$\Delta\mu=\Delta\mu_{\text{cpw}}\left(T_1\right)=-7\cdot10^{-2}$ (this
transition, manifested by a hysteresis S-loop in
$\Gamma\left(\Delta\mu\right)$, is outside the range chosen for figure
\ref{FigSecTrIs}).

As $\Delta\mu$ is further increased, the point representing the system in the
phase diagram approaches the shifted prewetting line,
$\Delta\tilde{\mu}_{\text{pw}}\left(T\right)$ (line 3), and the density
profiles begin to develop liquid-like fingers (see, e.g., figures
\ref{FigSecTrPr}(a) -- \ref{FigSecTrIs}(c)). The values of adsorption for the
fluid configurations with fingers belong to the branch of
$\Gamma\left(\Delta\mu\right)$, plotted in grey in figure
\ref{FigSecTrIs}(a). The values of the excess free energy belong to a
strictly concave branch of $\Omega^{\text{ex}}\left(\Delta\mu\right)$ (grey
curve in figure \ref{FigSecTrIs}(c)), signifying that, first, the fluid
states with liquid-like fingers are thermodynamically stable, i.e.
\emph{observable}, second, the configurations with liquid-like fingers form a
\emph{fluid phase}. As
$\Delta\mu\to\Delta\mu_2\equiv\Delta\tilde{\mu}_{\text{pw}}\left(T_1\right)$,
the length of the fingers increases and diverges in the limit (see again the
density profiles in figures \ref{FigSecTrPr}(a) -- \ref{FigSecTrIs}(c)).

To understand the phase with liquid-like fingers, we have analysed vertical
cross sections of various density profiles,
$\rho^{\text{cpd}}\left(x,y\right)$, belonging to that phase. We chose two
sets of values along the $x$-axis: $x_{\text{f}}$ -- between the capping wall
and the tips of the fingers, and $x_{\text{v}}$ -- well inside capillary
bulk, far away from the finger tips. We compared the slices
$\rho\left(x_{\text{f}},y\right)$ and $\rho\left(x_{\text{v}},y\right)$ with
the 1D density profiles of fluid configurations, coexisting inside the
associated slit pore during the shifted prewetting transition at
$\Delta\mu\equiv\Delta\mu_2=\Delta\tilde{\mu}_{\text{pw}}\left(T_1\right)$:
$\rho_{y,\text{film}}\left(y\right)$ and $\rho_{y,\text{vap}}\left(y\right)$.
We have found that an equivalence similar to the one expressed by equation
(27) of Part I, holds (within the margin of machine rounding error):
\begin{eqnarray}
\rho\left(x_{\text{f}},y\right)\equiv\rho_{y,\text{film}}\left(y\right),\nonumber\\ \rho\left(x_{\text{v}},y\right)\equiv\rho_{y,\text{vap}}\left(y\right).
\label{equalfing}
\end{eqnarray}

The above relation proves, that the profiles with liquid-like fingers
correspond to the \emph{continuous} onset of the thin film phase, which is
filling the capped capillary as $\Delta\mu\to\Delta\mu_2$, thus making
shifted prewetting a continuous phase transition.

The order parameter of the continuous prewetting is again adsorption,
$\Gamma\left(\Delta\mu\right)$, defined in equation (23) of Part I. While
$\Delta\mu<\Delta\mu_2$, the bulk of the capped capillary is in the vapour
phase (see figure \ref{FigSecTrIs}(b)), and the development and growth of the
denser phase from the capping wall leads to the divergence of the integral in
equation (23) of Part I, so that $\Gamma\to\infty$, as
$\Delta\mu\to\Delta\mu_2$. The value $\Delta\mu=\Delta\mu_2$ is the vertical
asymptote for the branch of $\Gamma\left(\Delta\mu\right)$-dependence (grey
line in figure \ref{FigSecTrIs}(a)). The branch of the excess free energy, on
the other hand, tends to a finite limit, as is typical in continuous phase
transitions (grey line in figure \ref{FigSecTrIs}(c)).

Increasing the control parameter, $\Delta\mu$, further above $\Delta\mu_2$
corresponds to a point, representing the system in the phase diagram,
advancing vertically between transition lines of shifted prewetting (line 3)
and CC (line 4), towards CC. For values of $\Delta\mu$ in the interval
$\Delta\mu_2\leq\Delta\mu\leq\Delta\mu_3$ the associated slit pore is in the
thin film phase (see figure \ref{FigSecTrIs}(b)). All thermodynamically
stable equilibrium fluid configurations in the capped capillary possess a
thin film at both side walls.

As the value of $\Delta\mu$ is increased further, the meniscus starts to
unbind from the capping wall into the capillary bulk (see density profiles in
figures \ref{FigSecTrPr}(d) -- \ref{FigSecTrPr}(f)). The analysis of vertical
cross sections (similar to the one presented above) shows that the structure
of the fluid between the capping wall of the capillary and the meniscus is
identical to that of capillary-liquid coexisting with the thin film in the
associated slit pore at $\Delta\mu=\Delta\mu_3$. The structure of the fluid
far from the capping wall and the meniscus has, in turn, been found to be
identical to the coexisting thin film phase. We also find the divergence of
the order parameter: $\Gamma\to\infty$, as $\Delta\mu\to\Delta\mu_3$; the
value $\Delta\mu=\Delta\mu_3$ provides the vertical asymptote for the
$\Gamma\left(\Delta\mu\right)$-dependence (black curve in figure
\ref{FigSecTrIs}(a)). Since in this limit the excess free energy tends to a
finite limit (black curve in figure \ref{FigSecTrIs}(c)), the unbinding
meniscus corresponds to the continuous CC.


We note that the density profiles, $\rho^{\text{cpd}}\left(x,y\right)$, from
the the Euler-Lagrange equation (equation (11) of Part I), can in fact be
obtained for values of $\Delta\mu$ in the larger interval
$\Delta\mu_1\leq\Delta\mu\leq\Delta\mu_3$, which includes the metastable part
of the thin film branch (the parts of black curves in figures
\ref{FigSecTrIs}(b) and \ref{FigSecTrIs}(c) between $\Delta\mu_1$ and
$\Delta\mu_2$). However, those states are thermodynamically unstable, which
follows immediately from the consideration of the
$\Gamma\left(\Delta\mu\right)$-dependence (see black curve in figure
\ref{FigSecTrIs}(a)). The part of the $\Gamma\left(\Delta\mu\right)$-branch
between $\Delta\mu_1$ and $\Delta\mu_2$ is non-monotonous, and, according to
the Gibbs rule (see equation (22) of part I), the corresponding part of the
excess free energy isotherm is not strictly concave. So, although the fluid
configurations corresponding to this part of the isotherm solve the
Euler-Lagrange equation, they do not minimise the free energy.

%

To summarise, we have shown that the planar prewetting transition can become
a continuous (second-order) phenomenon in a capillary-like geometry, when the
associated slit pore allows for the three-phase coexistence. Let us briefly
highlight the potential ramifications of this finding.

One important consequence is that it allows one to take prewetting to an
observable scale: the typical height of a thin film coexisting with vapour on
a planar surface is several molecular diameters, and rather involved and
specialised experimental techniques are required to even register an
existence of prewetting, see, e.g., reviews in references
\cite{HorikawaEtAlAdvCollSci11, HamraouiPrivatAdColIntSc09,
BrushiEtAlJChemPhys06, Bonn01}. However, the same phenomenon in a
capillary-like geometry corresponds to a singularity of an observable (the
adsorption, $\Gamma$), rather than to a tiny jump in its value, thus
registering a prewetting transition is simplified. Since the relation between
planar prewetting on a wall and shifted prewetting in the slit-like geometry
is clear (the value of the shift is constant for the entire transition line
and can be established experimentally by, e.g., calibration), using
capillary-like, rather than planar substrates can also simplify the
measurement of \emph{microscopic} prewetting transitions on various
substrates.

The continuous prewetting is related to the phenomenon of step wetting, where
a planar step of the height of several molecular diameters (the height should
be that of the coexisting thin film), immersed in vapour, adsorbs liquid,
showing a divergence of the meniscus as the prewetting pressure is approached
(consider, e.g. \cite{SaamJLowTempPhys09}). We again expect the substrate in
the form of a wide capped capillary to be more feasible to study
experimentally than a microscopically high step. For one, the height of the
step should be \emph{exactly} equal to the height of the coexisting thin film
(the value is very small: several molecular diameters). Using a capped
capillary completely eliminates the need to implement such a highly precise
microscopically patterned substrate: the continuous prewetting should be
observable at \emph{any} value of $H>H_3^{\text{cr}}$, where
$H_3^{\text{cr}}$ is the tri-critical value for the appearance of the
three-phase coexistence in a slit pore.

A further exploration of continuous prewetting would go beyond the scope of
the present work dedicated to wetting in capped capillaries. Of particular
interest would also be the study of critical exponents for the diverging thin
film fingers, as well as for the growing meniscus at $T>T_3^{\text{slt}}$.
Such investigation would allow us to establish Ising universality classes.
Another important direction is accounting for fluctuations, which may prevent
phase transitions from taking place in rather small systems at high
temperatures. We note, however, that increasing the wall separation, $H$,
would reduce the fluctuation effects allowing, in principle, for the
observation of the phenomena observed, which are not affected by increasing
$H$.

\section{Slow relaxation}
\label{SecDyn}

We have shown that reducing the attractive strength of the substrate
potential (lowering $T_{\text{w}}$) gradually leads to the appearance of a
metastable drop phase, which eventually stabilises for weaker substrates and
even coexists with vapour and capillary-liquid slab phases: see, e.g., the
upper isotherm in figure \ref{FigCritical}, where the concave branch of
excess free energy, defining the (metastable) drop phase (grey line), is
quite pronounced. We have restricted our consideration to fluids with vapour-
and liquid-like densities, so the temperatures were chosen above the bulk
triple point. At lower temperatures and/or weaker attractive substrates we
expect more metastable branches of free energy to appear, due to effects of
layering or freezing, see, e.g., \cite{BallEv}. In this section we use a
dynamic model to show how the relaxation of the system is affected by
metastability. In order to enhance the confinement effects of the substrate,
we deviate from the definition of the external potential, given in equation
(10) of Part I, and consider here a capped capillary formed from three 3-9
walls (see definition in equation (13) of Part I):
\begin{equation}
\label{DynSubs}
V\left(x,y\right)= V^{3-9}_{\varepsilon_{\text{w}},\sigma_{\text{w}}}\left(x\right)+V^{3-9}_{\varepsilon_{\text{w}},\sigma_{\text{w}}}\left(y\right)+V^{3-9}_{\varepsilon_{\text{w}},\sigma_{\text{w}}}\left(H-y\right).
\end{equation}

Using such a substrate potential leads to more pronounced hysteresis loops in
the adsorption isotherms. Substrates of the type defined by equation
\eref{DynSubs} were studied in detail in our previous work in reference
\cite{Yatsyshin2013}.

Under the standard assumptions of a local equilibrium \cite{Tarazona1} we use
``model A'' of the general dynamic universality class, describing the
evolution of an open system in contact with the reservoir of particles, which
keeps it at fixed values of $T$ and $\mu$ \cite{Hohenberg77,BookLang}:
\begin{equation}
    \label{dyndis}
    \frac{\partial\rho\left({\bf r},t\right)}{\partial t}=-\zeta\frac{\delta\Omega\left[\rho\right]}{\delta\rho\left({\bf r},t\right)},
\end{equation}
where $\zeta$ defines time units as
$\left(\zeta\varepsilon\sigma^3\right)^{-1}$. A particular value of $\zeta$
only affects the rate of relaxation to the equilibrium state without loss of
generality~\cite{Yatsyshin2012}. We then set $\zeta\equiv1$. The right hand
side of equation \eref{dyndis} is equivalent to the left hand side of the
Euler-Lagrange equation (see equation (11) of Part I). Thus, the stationary
solutions to the integral-differential equation \eref{dyndis} form the
extrema of fluid free energy and can be obtained independently by solving the
Euler-Lagrange equation. More applications of model \eref{dyndis} within the
DF framework can be found in, e.g., reference \cite{Th09}. Details of
numerical implementation of the implicit time-stepping scheme, which was used
to solve equation \eref{dyndis}, can be found in our previous work in
reference \cite{Yatsyshin2012}. A different model, incorporating the effects
of hydrodynamic interactions has recently been proposed in references
\cite{GoddardPRL12,GoddardJPhysCondMat13}.

\begin{figure}
\centering
    \includegraphics{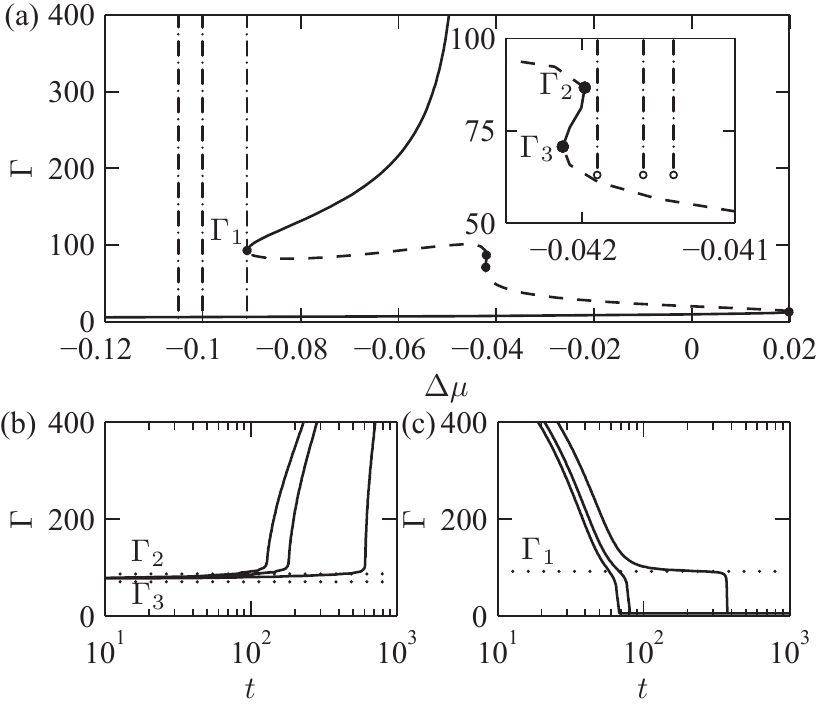}
    \caption{Evolution of the density profiles in the capped capillary,
    whose substrate potential is defined in equation \eref{DynSubs},
    with $\varepsilon_{\text{w}}=0.7$, $\sigma_{\text{w}}=2$, $H_0=5$, $H=30$; at $T=0.8T_{\text{c}}$, $\zeta=1$.
    (a) Equilibrium adsorption isotherm (bifurcation curve). Solid line:
    stable/metastable equilibria, dashed line: unstable states, vertical dashed-dotted lines:
    routes of dynamic emptying, starting with the same initial density profile of capillary filled
    with capillary-liquid at CC (at $\Delta\mu\equiv\Delta\mu_{\text{cc}}\left(T\right)=-4.65\cdot10^{-2}$),
    passing at distances $\left(\Delta\mu-\Delta\mu_1\right)$: $10^{-4}$, $9\cdot10^{-3}$ and $14\cdot10^{-3}$ to
    the spinodal at $\left(\Delta\mu_1,\Gamma_1\right)=\left(-0.091,92.5\right)$.
    Inset: vertical dashed-dotted lines show routes of dynamic filling starting with same initial profile of
    capillary filled with vapour, passing close to metastable branch, at the distances
    $\left(\Delta\mu-\Delta\mu_2\right)$: $8\cdot10^{-5}$, $4\cdot10^{-4}$, $6\cdot10^{-4}$ to its right-most
    spinodal at $\left(\Delta\mu_2,\Gamma_2\right)=\left(-0.041,86.7\right)$.
    (b) Evolution $\Gamma\left(t\right)$ for dynamic filling, note pinning to the metastable branch between
    $\left(\Delta\mu_2,\Gamma_2\right)$ and $\left(\Delta\mu_3,\Gamma_3\right) = 86.7$.
    (c) Evolution $\Gamma\left(t\right)$ for dynamic emptying; note the pinning to the spinodal at $\left(\Delta\mu_1,\Gamma_1\right)$.
    \label{FigFour}}
\end{figure}

In a dynamic setting, the adsorption isotherms act as bifurcation diagrams.
Figure \ref{FigFour}(a) shows the adsorption isotherm at $T=0.8$ for the
capillary with parameters $\varepsilon_{\text{w}}=0.7$,
$\sigma_{\text{w}}=2$, $H_0=5$, $H=30$. The fluid is treated in LDA, the
planar wetting temperature: $T_{\text{w}}=0.755$. The system exhibits only a
single first-order transition, namely the stable capillary prewetting
transition at $\Delta\mu_{\text{cpw}}\left(T\right)=-4.95\cdot10^{-2}$, but
there is also a metastable drop phase. Thermodynamically stable (and
metastable) fluid states form branches of the isotherm drawn with a solid
line, the unstable branches are drawn with a dashed line. The system has
three phases: vapour, corner drops and capillary-liquid slab. Their
corresponding fluid configurations form concave branches of excess free
energy. The capillary-liquid fills the entire capillary at
$\Delta\mu\equiv\Delta\mu_{\text{cc}}\left(T\right)=-4.65\cdot10^{-2}$

Choosing a density configuration, which minimises
$\Omega\left[\rho^{\text{cpd}}\left(x,y\right)\right]$ at some value of
$\mu\equiv\mu_{\text{sat}}+\Delta\mu$ as an initial condition, and then using
a different value in the equation \eref{dyndis}, allows to study the
relaxation of the system to a new equilibrium state. An independent
calculation of equilibria (steady solutions of equation \eref{dyndis}) was
used to control the convergence.

The resulting dynamics is most conveniently represented as evolution curves
for $\Gamma\left(t\right)$ (see figure \ref{FigFour}). We consider several
processes of dynamic capillary filling, where the initial condition belongs
to the vapour phase, and in the final state ($t\to\infty$) the entire
capillary is filled with capillary-liquid, figure \ref{FigFour}(b). We also
consider dynamic emptying, where the initial state of capillary-liquid
relaxes to vapour, figure \ref{FigFour}(c).

We find that the evolving system can spend considerable, albeit always
finite, time in the vicinity of metastable states. Figure \ref{FigFour}(b)
illustrates the pinning to a whole branch of metastable states, while figure
\ref{FigFour}(c) shows pinning to a single metastable state -- the spinodal
of the capillary prewetting.


\section*{Conclusion}

In this two-part study we have investigated the phonemenon of capillary
condensation. By applying a first-principles theory based on statistical
mechanics, we have modelled a prototypical 2D system, namely a slit pore
capped by an additional wall, and performed a full parametric study of the
model to show how the additional spatial dimension in the confined fluid
dramatically affects the nature of wetting. By manipulating the thermodynamic
fields acting on the system, it is possible to switch its wetting behaviour
between various mechanisms, or allow for several wetting mechanisms at the
same time. The planar wetting temperature, $T_{\text{w}}$, serves as a
convenient effective measure of the attractive strength exerted by the
substrate on the fluid.

Our study further demonstrates, how computational access to isotherms and the
concept of a Van der Waals loop can be used to analyse fluids in confiment.
Until quite recently only a handful of studies presented adsorption
isotherms, and even fewer presented complete phase diagrams of complex 2D
systems. We have explored all possible wetting scenarios in a capped
capillary, presented various phase diagrams, identifying every concave branch
of free energy with a fluid phase and revealed the physics of the interplay
between different wetting mechanisms. The numerical continuation technique
makes calculations readily accessible, and we expect this methodology to have
more applications in the mean-field DF calculations of fluids in a wide range
of settings. We note also that a consistent analysis of thermodynamic
stability of a given fluid configuration is impossible to carry out without
access to a set of density profiles, which allows to analyse the convexity of
the free energy. The arc-length continuation method allows for such analysis
in a natural and consistent way, as we have demonstrated. Although the most
technical details of continuation algorithms have been omitted, they are well
documented in our previous work, in reference~\cite{Yatsyshin2012}, in the
form, which makes them readily accessible for application to DF problems.

Apart from the general phase behaviour, we have also discussed in detail the
structure of coexisting fluid configurations, the excluded volume effects and
criticality. We now summarise the main results of this two-part study noting
that for the sake of clarity we also reiterate the main results of Part I:
\begin{itemize}
\item Capping a slit pore at one end by an additional wall leads to the
    \emph{capillary wetting transition}. It is a discontinuous
    first-order phase transition characterised by the presence of the
    \emph{capillary wetting temperature}, $T_{\text{cw}}$, which is a
    property of the pore and controls the order of CC. Capillary wetting
    can be manifested by, e.g., a continuous transition to CC at
    temperatures above $T_{\text{cw}}$, while below $T_{\text{cw}}$ CC
    remains abrupt (first-order).
\item There exists a first-order \emph{capillary prewetting} transition,
    which precedes a continuous CC. The capillary wetting transition can
    be viewed as the limiting case of the capillary prewetting at
    $T=T_{\text{cw}}$. In the $T$ -- $\Delta\mu$ space the transition
    line of capillary prewetting runs tangentially to the condensation
    line of capillary bulk. Capillary prewetting possesses a critical
    point at the \emph{critical capillary prewetting temperature},
    $T_{\text{cpw}}^{\text{cr}}$. However, we need to emphasise that the
    wetting on capped capillaries belongs to a different Ising
    universality class from that on a single planer wall. The critical
    exponents of an isothermal approach to capillary condensation above
    $T_{\text{cw}}$ can be found in references \cite{Yatsyshin2013,
    Parry07}.
\item Wetting on wedge-shaped substrates immersed in vapour is
    dramatically affected by the addition of a third wall, which changes
    the geometry to a capped capillary. On the phase diagram of the fluid
    in the $T$ -- $\Delta\mu$ space the remnant of wedge prewetting (see,
    e.g., reference \cite{RejDietNapPRE99}) can form a transition line,
    which is translated relatively to the wedge prewetting line along the
    $\Delta\mu$-axis by a constant value. This has been referred to as
    the \emph{shifted wedge prewetting} transition. The value of the
    shift depends on the width of the capillary and the parameters of the
    substrate forming it, with the mechanism of the shift being analogous
    to the shift of the bulk saturation inside a slit-like geometry
    (Kelvin shift). The relative value of the shift has been computed in
    one of the examples. We have also shown how the criticality
    associated with wedge prewetting affects the fluid configurations in
    the capillaries, whose phase diagram does not possess a transition
    line of the shifted wedge prewetting.
\item We have confirmed the existence of a triple point, where all three
    near-wall phases (vapour, drops and capillary-liquid slab), which are
    separately associated with shifted wedge prewetting and capillary
    prewetting, can coexist. Moreover, we have shown, that a tri-critical
    point must also exist.
\item The capillary bulk can exhibit a transition specific to a single
    planar wall, but shifted due to the presence of the additional wall
    -- shifted planar prewetting. It is well understood and known to be
    of first order. We have found a \emph{continuous prewetting}
    transition in capped capillaries, where the capillary bulk allows for
    shifted planar prewetting. In such systems the continuous CC is
    preceded by the continuous prewetting. In other words, the system has
    two consecutive continuous transitions, each associated with the
    divergence of adsorption. The adsorption isotherm of such capillaries
    consists of two unconnected diverging branches. We expect continuous
    prewetting to be easily accessible experimentally. It can potentially
    influence the experimental investigation of registering and measuring
    the characteristics of planar prewetting.
\item The metastable fluid configurations have been found to affect the
    evolution of the system, pinning it to spinodals of the corresponding
    phase branches of the free energy.
\end{itemize}

We now briefly highlight directions where further research is in order.
First, we note that in the presented density profiles the microscopic fluid
configurations have well defined angles between the two-phase interface
(level set at
$\left(\rho_{\text{cc}}^{\text{vap}}+\rho_{\text{cc}}^{\text{liq}}\right)/2$)
and the substrate see figures 2(d), 3, 5(a), 6(a), 8(c), 8(e) of part I and
figures \ref{FigVDSpr}(b), (c), \ref{FigTr2D} (d), (e) and \ref{FigSecTrPr}.
The value of the angle is \emph{different} from the contact angle at bulk
liquid -- vapour coexistence, which is often defined through Young's
equation. A deeper analytic exploration, possibly involving the concept of
line tension would allow to better understand that effect.

Considering further the presented density profiles, we note that there should
exist relations, connecting the values of the density profiles at contact
with the substrate with the thermodynamic fields present in the system ($T$,
$\mu$, $\varepsilon_{\text{w}}$, $\sigma_{\text{w}}$, $H$). Such relations
should exist for the contact at the corner apex ($\rho\left(x=0,y=0\right)$),
as well as at each capillary wall
($\rho^{\text{cpd}}\left(x\equiv0,y\right)$,
$\rho^{\text{cpd}}\left(x,y\equiv0\right)$).

Regarding the described phenomenology of wetting, we emphasise again that we
have not touched upon the universality of the reported phenomena. It is also
of great interest to obtain the critical exponents for the various
transitions we have presented, e.g., the power law describing
$\Delta\mu_{\text{cpw}}\left(T\right)$, as $T\to T_{\text{cw}}$, the power
law for the divergence of $\Gamma\left(\Delta\mu\right)$ as
$\Delta\mu\to\Delta\tilde{\mu}_{\text{pw}}\left(T\right)$ in the case of the
continuous prewetting transition. One might expect that the shifted wedge
prewetting belongs to a different universality class than the wedge
prewetting in the approach to bulk saturation, studied in, e.g., references
\cite{RejDietNapPRE99,ParryRasconWoodPRL99}.

An analytic expression for the shift of the wedge prewetting in a capped
capillary can be obtained following standard procedures, e.g.,
\cite{HendersonBook92}. The three-phase coexistence between vapour, drop and
capillary-liquid slab configurations requires additional investigation. We
have only reported on the existence of the triple line (as function of
parameter $H$), ending at the tri-critical point. The associated universality
class remains to be established.

Further, there might be a first-order transition, preceding the continuous
prewetting, similar to the way capillary prewetting precedes continuous
condensation. A discovery of such a phase transition (or a proof that it does
not exist) would then complete the picture of condensation in slit-like 2D
geometries. Finally of interest would be how the various phase transitions
are influenced by chemical and/or topographical wall heterogeneities which
are known to have a substantial influence on wetting,
e.g.~\cite{Savva2010,Vellingiri2011}. We hope to investigate these and
related issues in future studies.

\section*{Acknowledgements}
We are grateful to Prof. S. Dietrich for a fruitful discussion on wedge
wetting and to the European Research Council via Advanced Grant No. 247031
for support of this research.

\section*{References}

\end{document}